\documentstyle[amssymb,preprint,pra,aps]{revtex}

\begin{document}
\draft
\title{{\large Decoherence of a Pointer by a Gas Reservoir}}
\author{{Michael Kleckner}$\thanks{%
Corresponding author: kleckner@tx.technion.ac.il}${\ and Amiram Ron }}
\address{Department of Physics, \\
{Technion - Israel Institute of Technology}, \\
Haifa 32000, Israel}
\date{\today}
\maketitle

\begin{abstract}
\label{abst}

We study the effect of the environment on the process of the measurement of
a state of a microscopic spin half system. The measuring apparatus is a
heavy particle, whose center of mass coordinates can be considered at the
end of the measurement as approximately classical, and thus can be used as a
pointer. The state of the pointer, which is the result of its interaction
with the spin, is transformed into a mixed state by the coupling of the
pointer to the environment. The environment is considered to be a gas
reservoir, whose particles interact with the pointer. This results in a
Fokker-Planck equation for the reduced density matrix of the pointer. The
solution of the equation shows that the quantum coherences, which are
characteristic to the entangled state between the probabilities to find the
pointer in one of two positions, decays exponentially fast in time. We
calculate the exponential decay function of this decoherence effect, and
express it in terms of the parameters of the model.
\end{abstract}


\newpage

\narrowtext

\section{\bf Introduction}

In recent years there has been a considerable interest in the effect of the 
{\em environment }on the process of {\em quantum measurement}. The
measurement of physical variable of a {\em microscopic} system - a system
which is described by quantum theory - is usually realized by coupling the
system to a {\em ''classical meter''}. The meter itself does not have to be
macroscopic, however some of its physical variables, say the position of its
center of mass, can be considered as approximately {\em classical}. If these
variables are chosen to be correlated with those of the system, their
readings are related to the measurement of the microscopic properties of the
system. When the measurement process is efficient there is one to one
correspondence between the macroscopic readings of the meter and the
microscopic quantities, which are set to be measured. A particle, whose
center of mass position serves as a meter, will be identified as a {\em %
pointer}. A theory that starts from a model Hamiltonian, which includes only
the system, the pointer, and the coupling between them, cannot account for
the real measurement process. The reason is that the classical meter is
significantly influenced, on the time scale of its motion, by the
environment around it. Even though the dynamics of the combined entity - the
system, the meter, and the environment - can be described by quantum
mechanics, the environment ({\em bath}) plays a distinctive role in the
process.

The role of the environment is{\em \ formally} expressed in the quantum
description of the experiment in that the system ends up in a {\em mixed
state}, rather then in a {\em pure state,} which is the outcome of a
''pure'' unitary transformation. While a unitary transformation leaves the
entire entity, system, meter and bath, in an{\em \ entangled state, }so that
the pointer's final positions are not definite, but rather display {\em %
interference,} the environment generates{\em \ decoherence,} and induces
definite final readings of the meter. The fundamental nature of the quantum
theory makes the predictions of the readings {\em probabilistic}, while the
decoherence, which is caused by the nature of the environment, is
responsible for the {\em definite} positions of the pointer.

That decoherence is induced by the environment was recognized long ago, as
was pointed out explicitly almost twenty years ago by Zurek \cite{Zurek
82,Zurek 91,Zurek 97}. This point was previously noticed by Zeh \cite{Zeh}
and K\"{u}bler and Zeh \cite{Kubler 73} in the early seventies. Soon later
Caldeira and Leggett \cite{Caldeira 85,Caldeira 83}, and Walls, Milburn and
Collet \cite{Milburn 85,Collet 85} have shown, using a harmonic oscillator
as their microscopic quantum system, to be {\em measured}, and an ensemble
of many harmonic oscillators to model the environment, that indeed the
environment introduces decoherence to the Density Matrix of the quantum
system. In Ref. \cite{Caldeira 85} an initial pure state of two spatially
separated Gaussian wave-packets of the microscopic system - a harmonic
oscillator - are turned by the bath, in an extremely short time, into a well
defined mixed state of the two wave-packets. Ref. \cite{Milburn 85} starts
from a pure state with a superposition of two separated pure coherent states
of a harmonic oscillator, and ends up, again due to the interaction with the
bath, in a mixed state of the two coherent states. The quantum interferences
between the two wave-packets, or the two coherent states, fade away in an
extremely short time. K\"{u}bler and Zeh \cite{Kubler 73} and later Zurek 
{\it et al. }\cite{Zurek 93} have shown that the coherent states of a
harmonic oscillator in a bath are somewhat unique in their decoherence,
being almost classical. Collett \cite{Collet 88} has extended the
investigation of decoherence using density-matrix calculations for some
simple {\em open systems}. Following the ingenious experiments of the Ecole
Normal Superieure Group \cite{Brune 96} Paz and Zurek \cite{Paz 99} have
investigated the decoherence in the limit of weak interaction between the
''meter'' and the bath. Recently the decoherence of a harmonic oscillator,
playing the role of a meter measuring a spin half system, has been studied \
by Venugopalan \cite{Venugopalan 00}. Their model for the bath was the
standard one \cite{Caldeira 85}, namely an ensemble of harmonic oscillators.

In the present paper we revisit the decoherence problem, and study it more
in the context the {\em ''measurement'' theory} in a somewhat different
setup. We consider a simple model to describe a combined system, which is
made of three coupled parts, a microscopic atomic system, a measuring
apparatus, and a bath. While the standard model for the bath is an ensemble
of independent harmonic oscillators, we consider here a {\em gas reservoir},
namely an ensemble of independent particles, which interact with the meter.
The initial state of a {\em spin-half} atom is set to be {\em ''measured''}%
{\it \ } by the final position of a {\em massive particle}, which is, in
turn, in contact with the {\em reservoir}. This combined system is supposed
to simulate the entire process of a {\em ''real''} measurement. An
environment of a gas seems to be more appropriate for the description of the
motion of a particle-pointer. Furthermore, by changing the density of the
gas one can control the time of decoherence. The atom interacts for a very
short time with the meter, and the state of the two systems is becoming {\em %
entangled}. The two macroscopically separated positions of the pointer,
which are now quantum-mechanically correlated with the initial state of the
spin, {\em interfere} with each other. The interaction of the meter with the
particles of the bath introduces decoherence, namely is destroying the
interference, and making the pointer's readings distinguishable. The
development in time of the entire system is described by a unitary
transformation, which originates probability into the dynamics by quantum
mechanics. This evolution in time is conveniently described by the {\em %
density-matrix} of the entire system. In the treatment of the measurement
process, a pure state is transformed after integrating out the environment
into a mixed state, and the notion of the {\em collapse} of the wave
function is never being introduced. The probabilistic nature of microscopic
dynamics, which is built into the quantum description, is expressed, at the
end of the process, in terms of the two {\em potentially definite} positions
of the pointer.

The crucial role of the environment is clearly expressed in the stage where
the density-matrix of the combined system of the pointer and the bath is 
{\em traced over} the degrees of freedom of the gas particles of the
reservoir. Since the exact initial conditions of the bath cannot be
completely prescribed, the unitary transformation, which controls the motion
in time, does not account for the {\em complete} description of the entire
process. The point is that even if we could theoretically prescribe the
state of the bath at a given time, it cannot be completely decisive for an
actual experiment. The reason is that the rate of change of the environment
states is order of magnitudes faster than the inverse time scales of the
experiment. Since these time scales in reality are so different, even the
notion of {\em initial conditions} is not well defined. Relevant initial
conditions can be meaningfully prescribed, for two interacting systems, only
when their intrinsic time scales are not orders of magnitudes apart. A
theory, which is being introduced to study a real experiment, has to
incorporate these facts into its structure. It is manifested here by
assuming that the reservoir is in {\em thermal equilibrium}. This has
nothing to do with the consistency of quantum theory, but rather with the
role of the real initial conditions of the entity we set to investigate. It
is our opinion that a theory, being a logical construction in terms of a
mathematical scheme, is supposed to be correlated with the {\em intrinsic
approximate nature} of the results of real experiments, and it has to
reflect this facet of physical reality.

The outline of the paper is as follows: In Section II we introduce the model
to be investigated in terms of the Hamiltonians for the spin-half atom, the
massive pointer, the gas reservoir, and the interactions among them. Section
III is concerned with the motion of the atom and the pointer without the
reservoir. The quantum interference between the two final positions of the
pointer is studied in detail. In Section IV we add the bath into the
development in time, introduce the {\em Fokker-Planck} equation for the
pointer in the gas environment, and show how the decoherence is set in. In
Section V, we discuss the conclusions. The details of the solution of the
Fokker-Planck equation for the {\em Wigner} {\em distribution} is outlined
in Appendix A. In Appendix B we study an intuitive model for the bath, which
is simulated by a {\em random field} perturbing the meter, and compare it
with the results obtained for the gas reservoir.

\section{Model Hamiltonian}

We consider the {\em measurement} of the spin components of a{\em \ spin half%
} atom. We introduce, for our {\em microscopic system} $S,$ i.e., the {\em %
spin half }atom, the Hamiltonian, $H_{S},$ which, in the absence of an
external magnetic field, has degenerate energy eigenstates

\begin{equation}
H_{S}\left| \pm 1\right\rangle =\epsilon _{0}\left| \pm 1\right\rangle .
\label{1.1}
\end{equation}
These are also eigenstates of $\sigma _{z}$, the component of the spin
operator along the $z$-axis, i.e., $\sigma _{z}\mid \sigma \rangle =\sigma
\mid \sigma \rangle ,$ where $\sigma =\pm 1,$ for the {\em up} or {\em down}
states. We shall take $\epsilon _{0}=0$.

Our measuring device, the{\it \ }{\em pointer}, is taken to be a {\em massive%
} particle of mass $M$ (possibly a heavy atom), whose {\em center of mass}
position and momentum are ${\bf R}\ $and$\ {\bf P}$ respectively. The
Hamiltonian, $H_{P}$, of the pointer is then

\begin{equation}
H_{P}=\frac{P^{2}}{{\ }2M}.  \label{1.2}
\end{equation}
The pointer is the {\em meter} which is assigned to perform a measurement of
the $z-$components of the spin of the atom. A {\em measurement} is thought
of as a process, which generates a correlation between the measured property
of the microscopic system, i.e. the spin component, and say, the final
macroscopic position of the pointer. To measure the spin variable $\sigma
_{z}$, which has the observable outcomes $\pm 1$, we adopt, after Peres \cite
{Peres 89}, the following {\em coupling} Hamiltonian:

\begin{equation}
H_{SP}=V(t)P_{x}\sigma _{z},  \label{1.3}
\end{equation}

\noindent where $V(t)$ is a c-number time dependent ''impact'' function,
which couples the spin operator to the pointer. The function $V(t)$ has the
dimension of velocity, and is active for a very short finite interaction
time, $0\leq t\leq T$. We wish to point out that one massive atom can play
the role of both the pointer and the spin half system, when the internal
microscopic variable - the spin - is being coupled for a short time with the
center of mass of the atom, as in a Stern-Gerlach device for example.

The effect of the environment on the pointer is to be studied here assuming
the following picture. The pointer is considered to be immersed in a bath,
which is viewed as a set of {\em independent identical ''field'' particles, }%
whose $i$-th particle is described by the mass $m,$ coordinate\ ${\bf r}%
_{i},\ $and momentum ${\bf p}_{i},$ and the Hamiltonian 
\begin{equation}
H_{B}=\sum_{i}\frac{p_{i}^{2}}{2m}.  \label{1.4}
\end{equation}
Each of these bath particles interacts with the pointer via a potential $%
\phi ({\bf R}-{\bf r}_{i}),$ and thus the pointer - bath interaction
Hamiltonian is 
\begin{equation}
H_{PB}=\ \sum_{i}\phi ({\bf R}-{\bf r}_{i}).  \label{1.5}
\end{equation}
The total Hamiltonian for our system is thus 
\begin{equation}
H=H_{S}+H_{P}+H_{SP}+H_{B}+H_{PB}.  \label{1.6}
\end{equation}

The outline of the procedure employed for the measurement of the spin of the
atom is as follows. Suppose that the atom is initially in the state

\begin{equation}
|\psi _{S}(0)\rangle =\sum_{\sigma }a_{\sigma }|\sigma \rangle =a_{+1}\left|
+1\right\rangle +a_{-1}\left| -1\right\rangle ,  \label{1.7}
\end{equation}

\noindent where $\ a_{\sigma }$ is the amplitude to find the spin in the $%
\sigma $ state, $a_{\sigma }^{\ast }$ is the complex conjugate of $a_{\sigma
},$ and $\sum_{\sigma }a_{\sigma }^{\ast }a_{\sigma }=1$. This state is to
be measured by the position of the pointer after it interacts with the atom.
The initial state of the pointer, which is also described by quantum
mechanics, is assumed to be a {\em Gaussian wave packet}:

\begin{equation}
\psi ({\bf R},0)=(2\pi \Delta ^{2})^{-3/4}\ e^{-R^{2}/4\Delta ^{2}},
\label{1.8}
\end{equation}

\noindent where $\Delta $ is the {\em spread} of the wave packet in $R$,
namely, the initial expectation value of $R$ is $0$, and that of $R^{2}$ is $%
\Delta ^{2}$. We take $\Delta $ to be as small as possible, to have a better
resolution of the position of the pointer. The change of the state of the
pointer in time, due to its interaction with the atom and the environment,
is now studied in details, by solving the equation of motion of the pointer
and {\em eliminating} the degrees of freedom of the ''gas'' of field
particles surrounding the pointer. But before we will do that, let us first
take a look at the case where there is no bath and the measurement of the
spin is attempted with a ''free'' pointer.

\section{Measuring the Spin by a free Pointer}

First we consider our system without the environment, and find the time
development of the pointer from its given initial conditions. Since the
''free'' spin term of the atom is irrelevant, the Hamiltonian is

\begin{equation}
H=\frac{{\bf P}^{2}}{2M}+V(t)P_{x}\sigma _{z}.  \label{2.1}
\end{equation}
The state of the combined system, atom and pointer, can be given at the time 
$t$, by the wave function $\Psi ({\bf R},\sigma ,t),$ which is the solution
of the Schr\"{o}dinger's equation

\begin{equation}
i\hbar {\dot{\Psi}}({\bf R},\sigma ,t)=\left( -\frac{{\hbar ^{2}}}{2M}\frac{%
\partial ^{2}}{\partial {\bf R}^{2}}\ -i\hbar \sigma V(t){\frac{\partial }{{%
\ \partial X}}}\right) \Psi ({\bf R},\sigma ,t),  \label{2.2}
\end{equation}
with the initial condition

\begin{equation}
\Psi ({\bf R},\sigma ,0)=a_{\sigma }\ (2\pi \Delta ^{2})^{-3/4}\
e^{-R^{2}/4\Delta ^{2}}.  \label{2.3}
\end{equation}
It is convenient to introduce the momentum representation for the pointer,
namely ${\bf P}\left| {\bf k}\right\rangle =\hbar {\bf k}\left| {\bf k}%
\right\rangle ,$ and apply the {\em Fourier transform} {\em \ }to the wave
function, 
\begin{equation}
\Psi ({\bf R},\sigma ,t)={\frac{1}{({2\pi )}^{3}}}\int_{-\infty }^{+\infty
}\ d^{3}k\ e^{+i{\bf k\cdot }{\bf R}}\Psi ({\bf k},\sigma ,t).  \label{2.4}
\end{equation}
The equation of motion takes the form 
\begin{equation}
i\hbar \frac{\partial }{\partial t}\Psi ({\bf k},\sigma ,t)=\left( \frac{%
\hbar ^{2}k^{2}}{2M}+V(t)\hbar k_{x}\sigma \right) \Psi ({\bf k},\sigma ,t),
\label{2.5}
\end{equation}
where $k=\left| {\bf k}\right| ,$ and the initial condition is given by

\begin{equation}
\Psi ({\bf k},\sigma ,0)=a_{\sigma }(8\pi \Delta ^{2})^{3/4}\ e^{-\Delta
^{2}k^{2}}.  \label{2.6}
\end{equation}

\noindent The solution of Eq.~(\ref{2.5}) is then

\begin{equation}
\Psi ({\bf k},\sigma ,t)=\Psi ({\bf k},\sigma ,0)e^{-i\omega _{k}t-i%
\overline{X}(t)\ k_{x}\sigma },  \label{2.7}
\end{equation}
where \ $\omega _{k}=\frac{\hbar k^{2}}{2M},$ and \ $\overline{X}%
(t)=\int_{0}^{t}\ dt^{\prime }V(t^{\prime }).$

It is convenient to discuss the measurements in terms of the{\em \ density
matrix} of the combined system. In the $({\bf k},\sigma )$ representation we
write the density matrix as

\begin{equation}
\rho ({\bf k},\sigma ;{\bf k}^{\prime },\sigma ^{\prime };t)=\Psi ^{*}({\bf %
k^{\prime }},\sigma ^{\prime },t)\ \Psi ({\bf k},\sigma ,t),  \label{2.8}
\end{equation}
where $\Psi ^{*}$ is the complex conjugate of $\Psi $. From Eq.~(\ref{2.7})
we get

\begin{eqnarray}
\rho ({\bf k},\sigma ;{\bf k}^{\prime },\sigma ^{\prime };t) &=&\rho ({\bf k}%
,\sigma ;{\bf k}^{\prime },\sigma ^{\prime };0)  \nonumber \\
&&\times e^{-i(\omega _{k}-\omega _{k^{\prime }})t}e^{-i\overline{X}(t)\
(k_{x}\sigma -k_{x}^{\prime }\sigma ^{\prime })},  \label{2.9}
\end{eqnarray}
where, using Eq.~(\ref{2.6}), the density matrix is initially

\begin{equation}
\rho ({\bf k},\sigma ;{\bf k}^{\prime },\sigma ^{\prime };0)=a_{\sigma
^{\prime }}^{*}a_{\sigma }\ (8\pi \Delta ^{2})^{3/2}\ e^{-\Delta
^{2}(k^{2}+k^{\prime }{}^{2})}.  \label{2.10}
\end{equation}
We now return to real space, and transform the density matrix to the $({\bf R%
},\sigma )$ representation \noindent to find:

\widetext 

\begin{eqnarray}
\rho ({\bf R},\sigma ;{\bf R}^{\prime },\sigma ^{\prime };t) &=&a_{\sigma
^{\prime }}^{*}a_{\sigma }(2\pi \Delta ^{2}\xi (t))^{-3/2}  \nonumber \\
&&\times \exp \left( -\frac{1}{{4\Delta ^{2}}\xi (t)}\left( {\zeta ^{*}}(t)[%
{\bf R}-{\hat{{\bf x}}}\overline{X}(t)\sigma ]^{2}+{\zeta }(t)[{\bf R}%
^{\prime }-{\hat{{\bf x}}}\overline{X}(t)\sigma ^{\prime }]^{2}\right)
\right) ,  \label{2.11}
\end{eqnarray}
where $\zeta (t)=1+it/\tau _{f},$ $\xi (t)=\zeta ^{*}(t)\zeta (t)=1+(t/\tau
_{f})^{2}.$ Here $\tau _{f}=2M\Delta ^{2}/\hbar $ \ is a characteristic time
of {\em free quantum diffusion} of the pointer, and ${\hat{{\bf x}}}$ is a
unit vector along the $x-$ axis. Eq.~(\ref{2.11}) is our result for the
density matrix of the ''free'' pointer. In passing we notice that we can
express formally the density matrix at time $t$ in terms of the density
matrix at $t=0$, namely

\begin{equation}
\rho ({\bf R},\sigma ;{\bf R}^{\prime },\sigma ^{\prime };t)=\int_{-\infty
}^{+\infty }\ d^{3}Y\ \int_{-\infty }^{+\infty }\ d^{3}Y^{\prime }\ J({\bf R}%
,{\bf R}^{\prime },t;{\bf Y},{\bf Y}^{\prime },0)\ \rho ({\bf Y},\sigma ;%
{\bf Y}^{\prime },\sigma ^{\prime };0),  \label{2.12}
\end{equation}

\noindent where the {\em propagation kernel} is

\begin{eqnarray}
J({\bf R},{\bf R}^{\prime },t;{\bf Y},{\bf Y}^{\prime },0) &=&{\frac{1}{{%
(2\pi )^{6}}}}\int_{-\infty }^{+\infty }\ d^{3}k^{\prime }\ e^{-i{\bf k}%
^{\prime }\cdot ({\bf R}^{\prime }-{\bf Y}^{\prime })}\int_{-\infty
}^{+\infty }\ d^{3}k\ e^{i{\bf k\cdot }({\bf R}-{\bf Y})}e^{-i(\omega
_{k}-\omega _{k^{\prime }})t-i\overline{X}(t)\ (k_{x}\sigma -k_{x}^{\prime
}\sigma ^{\prime })}  \nonumber \\
&=&\left( \frac{M}{2\pi \hbar t}\right) ^{3}\left[ e^{iM\left[ {\bf R}-{\bf %
Y-}{\hat{{\bf x}}}\overline{X}(t)\sigma \right] ^{2}/2\hbar t}e^{-iM\left[ 
{\bf R}^{\prime }-{\bf Y}^{\prime }-{\hat{{\bf x}}}\overline{X}(t)\sigma
^{\prime }\right] ^{2}/2\hbar t}\right] .  \label{2.13}
\end{eqnarray}
\narrowtext

We also wish to point out that we could have obtained Eq.~(\ref{2.11})
directly from Eq.~(\ref{2.7}) using

\begin{eqnarray}
\Psi ({\bf R},\sigma ,t) &=&{\frac{1}{{(2\pi )^{3}}}}\int_{-\infty
}^{+\infty }\ d^{3}k\ e^{-i{\bf k}^{\prime }{\bf R}^{\prime }}\Psi ({\bf k}%
,\sigma ,t)  \nonumber \\
&=&a_{\sigma }(2\pi \Delta ^{2}\zeta ^{2}(t))^{-3/4}  \nonumber \\
&&\times \exp \left( -\frac{1}{4\Delta ^{2}\zeta (t)}[{\bf R}-{\hat{{\bf x}}}%
\overline{X}(t)\sigma ]^{2}\right) ,  \label{2.14}
\end{eqnarray}
and $\rho ({\bf R},\sigma ;{\bf R}^{\prime },\sigma ^{\prime };t)=\Psi ^{*}(%
{\bf R}^{\prime },\sigma ^{\prime },t)\Psi ({\bf R},\sigma ,t).$

We now turn to study Eq.~(\ref{2.11}). First we observe that our density
matrix represents a {\em pure state}. It is just the evolution in time, by a 
{\em unitary transformation }\ of the initial {\em product state, }$\Psi
(0)=(a_{+1}+a_{-1})\psi ({\bf R},0),$ into an {\em entangled state} of the
particle and the pointer, namely

\begin{equation}
\Psi (t)=a_{+1}\psi _{+}({\bf R,}t)+a_{-1}\psi _{-}({\bf R},t),  \label{2.15}
\end{equation}
where

\begin{eqnarray}
\psi _{\pm }({\bf R}) &=&(2\pi \Delta ^{2}\zeta ^{2}(t))^{-3/4}  \nonumber \\
&&\times \exp \left( -\frac{1}{4\Delta ^{2}\zeta (t)}[{\bf R}\mp {\hat{{\bf x%
}}}\overline{X}(t)]^{2}\right) ,  \label{2.16}
\end{eqnarray}
are two displaced wave packets of the pointer. In this way the amplitude of
the position of the pointer is correlated to the spin of the particle. This
is not yet a measurement, but rather an establishment of quantum mechanical
correlations, due to the interaction between the ''free'' pointer and the
particle.

The outcome of a {\em measurement} of the position of the pointer is then 
{\em predicted}, using the combined system's density matrix, by calculating
the {\em probability density} to detect the center of mass of the pointer
around ${\bf R}$, i.e., $P({\bf R},t)=\sum_{\sigma ,\sigma ^{\prime }}\rho (%
{\bf R},\sigma ;{\bf R},\sigma ^{\prime };t).$ Since the measurement is done
after the interaction was completed, namely at a time $t>T$, we set $%
\overline{X}(t)=\overline{X}=\int_{0}^{T}\ dt^{\prime }V(t^{\prime }),$ and
write $P({\bf R},t)$ as a sum of three terms:

\begin{equation}
P({\bf R},t)=P_{+1}({\bf R},t)+P_{-1}({\bf R},t)+P_{+1,-1}({\bf R},t).
\label{2.17}
\end{equation}

\noindent Using $\Delta _{f}^{2}(t)=\Delta ^{2}\xi (t),$\ \ the first term,
with \ $\sigma =\sigma ^{\prime }=+1,$ is 
\begin{eqnarray}
P_{+1}({\bf R},t) &=&\mid a_{+1}\mid ^{2}(2\pi \Delta _{f}^{2}(t))^{-3/2} 
\nonumber \\
&&\times \exp \left( -\frac{1}{2\Delta _{f}^{2}(t)}[{\bf R}-{\hat{{\bf x}}}%
\overline{X}]^{2}\right) ,  \label{2.18a}
\end{eqnarray}
and second term, with \ $\sigma =\sigma ^{\prime }=-1,$ is 
\begin{eqnarray}
P_{-1}({\bf R},t) &=&\mid a_{-1}\mid ^{2}(2\pi \Delta _{f}^{2}(t))^{-3/2} 
\nonumber \\
&&\times \exp \left( -\frac{1}{2\Delta _{f}^{2}(t)}[{\bf R}+{\hat{{\bf x}}}%
\overline{X}]^{2}\right) ,  \label{2.18b}
\end{eqnarray}
corresponding to the probabilities of up and down spins. While the third
term, with $a_{\pm 1}=\left| a_{\pm 1}\right| e^{i\varphi _{\pm }},$ is

\begin{eqnarray}
P_{+1,-1}({\bf R},t) &=&2\sqrt{P_{+}({\bf R},t)P_{-}({\bf R},t)}  \nonumber
\\
&&\times \cos \left( \frac{\overline{X}X}{\Delta _{f}^{2}(t)}\frac{t}{\tau
_{f}}+\varphi _{-}-\varphi _{+}\right) ,  \label{2.19}
\end{eqnarray}
is due to {\em quantum interference} between the two spin states. In Eqs.~(%
\ref{2.18a}), (\ref{2.18b}) and (\ref{2.19}) 
\begin{equation}
\Delta _{f}^{2}(t)=\Delta ^{2}[1+(t/\tau _{f})^{2}]  \label{2.20}
\end{equation}
$\ $ reflects the spatial broadening of the pointer's position as a result
of the free quantum diffusion. At this point we note that${\rm \ }$if the
''pointer'' is a Silver atom, with a mass of $M=1.8\times 10^{-22}$ {\rm g},
and initial spread of position $\Delta =1$ ${\rm \mu }${\rm m},

\begin{equation}
\tau _{f}={\frac{{2M{\Delta }^{2}}}{\hbar }}\simeq 3\text{ }{\rm ms}
\label{2.21}
\end{equation}

\noindent which is a long time on the scale of a ballistic experiment of an
atom.

The probability distribution for the position of the pointer, long after it
departed from the ''measured'' particle, has three components: two of them
are ''normal'', and the third one is ''strange''. As it is clear from Eqs.~(%
\ref{2.18a}) and (\ref{2.18b}), one component represents a positive
deflection of the pointer by $+\overline{X},$ and displays a Gaussian
centered around $(+\overline{X},0,0),$ which is correlated with the {\em %
spin up} state. The other one indicates a negative deflection by $-\overline{%
X},$ and displays a Gaussian centered around $(-\overline{X},0,0),$ which is
correlated with the {\em spin down} state of the measured system. The width
of each term is of the order of the spread of the initial distribution.
However the weight of each peek, i.e., the probability of positive or
negative deflection, is decided by the initial probability of the two spin
states, namely by $\mid a_{\sigma }\mid ^{2}$. The probability distribution
of Eq.~(\ref{2.19}) exhibits an interference between the two displaced
Gaussians. Notice that this interference term oscillates in time with a
frequency given by

\begin{equation}
\Omega _{int}=\frac{\overline{X}X}{\Delta ^{2}}\frac{1}{\tau _{f}}.
\label{2.22}
\end{equation}
This interference is an unambiguous outcome of the quantum mechanical
development in time of the pure entangled state of the pointer and the
system to be measured. This result is completely {\em reversible} in time,
and in principle can be reversed to the initial setup.

\section{The Pointer in a Bath}

We turn now to consider the case in which, in addition to the ''external''
interaction with the spin, the pointer is coupled to a reservoir. Our task
is to describe the evolution in time of the pointer's state under the
influence of this reservoir. To this end, we will drive a master equation
for the {\em reduced density matrix} $\rho (t)$ of the pointer, following
the method outlined in Ref. \cite{CCT 92}. The main ingredients of this
method are as follows: The effect of the pointer on the bath is assumed to
be very small, and the interaction $H_{PB}$ is considered as a weak
perturbation. The equation of motion for the density matrix of the combined
system, pointer and bath, is written up to second order in $H_{PB}.$
Finally, by tracing over the degrees of freedom of the bath, one obtains an
equation of motion for the pointer only, which is valid on time scales much
larger compared with the typical correlation times of the bath. We choose
here a particular model for the bath, which is constituted of a set of
independent field particles, and trace over the dynamical variables of these
particles. This results in a Fokker - Planck equation for the reduced
density matrix, which can then be solved for the time dependent behavior of
the pointer.

\subsection{\sl The Reduced Density Matrix}

In order to study the effect of the interaction, $H_{PB},$ on the pointer,
it will be convenient to introduce the Fourier transform of the interaction
potential, $\phi ({\bf r}),$ in space, in a volume $V$: 
\begin{equation}
\phi ({\bf r})=\frac{1}{V}\sum_{{\bf q}}e^{i{\bf q}\cdot {\bf r}}\ \phi (%
{\bf q}),  \label{3.1}
\end{equation}
Then the interaction Hamiltonian can be written as

\begin{equation}
H_{PB}={\frac{1}{V}}\sum_{{\bf q}}e^{i{\bf q}\cdot {\bf R}}\ \phi ({\bf q})\
n({\bf q}),  \label{3.2}
\end{equation}
where the Fourier transform of the {\em density }\ (operator) of the field
particles is 
\begin{equation}
n({\bf q})=\sum_{i}e^{-i{\bf q}\cdot {\bf r}_{i}}.  \label{3.3}
\end{equation}

To get the equation of motion for the reduced density matrix for the
pointer, we transform to the {\em Interaction Representation (IR),} and
write the Hamiltonian as

\begin{equation}
H_{PB}(t)={\frac{1}{V}}\sum_{{\bf q}}\ \phi ({\bf q})\ n({\bf q},t)\ A_{{\bf %
q}}(t),  \label{3.4}
\end{equation}
where, for the pointer 
\begin{equation}
A_{{\bf q}}(t)=e^{iH_{p}t/\hbar }\ e^{i{\bf q}\cdot {\bf R}}\
e^{-iH_{p}t/\hbar },  \label{3.5}
\end{equation}
and for the bath operator

\begin{equation}
\ n({\bf q},t)=e^{iH_{B}t/\hbar }n({\bf q})\ e^{-iH_{B}t/\hbar }.
\label{3.6}
\end{equation}

\noindent The reduced density matrix of the pointer, in the IR,

\begin{equation}
\overline{\rho }(t)=Tr_{B}\left\{ e^{i(H_{P}+H_{B})t/\hbar }\ {\rho }%
(t)e^{-i(H_{P}+H_{B})t/\hbar }\right\} ,  \label{3.7}
\end{equation}

\noindent obeys the following equation, \cite{CCT 92} Eq.~(B.30):

\begin{eqnarray}
\frac{\Delta \overline{\rho }(t)}{\Delta t} &=&-{\frac{1}{{\hbar ^{2}}}}{%
\frac{1}{{\Delta t}}}\int_{t}^{t+\Delta t}\ dt_{1}\int_{t}^{t_{1}}\ dt_{2}{%
\frac{1}{V}}\sum_{{\bf q}}\ \phi ({\bf q}){\frac{1}{V}}\sum_{{\bf q^{\prime }%
}}\ \phi ({\bf q^{\prime }})  \nonumber \\
&\ \ &\ Tr_{b}\{[\ n({\bf q},t_{1})\ A_{{\bf q}}(t_{1}),[\ n({\bf q^{\prime }%
},t_{2})\ A_{{\bf q^{\prime }}}(t_{2}),\ \overline{\rho }(t)\ \otimes \ \rho
_{B}]\},  \label{3.8}
\end{eqnarray}

\noindent where the{\em \ trace} is taken over the field particles' degrees
of freedom, and $\rho _{B}$ is the reduced density matrix of the bath. This
master equation, Eq. (\ref{3.8}) is similar to Eq.(1) of Paz and Zurek \cite
{Paz 99}, and to that of Unruh and Zurek \cite{Unruh 89}; however unlike the
scalar potential that simulates their bath, here the environment is a gas.
Another general master equation was derived by Joos \cite{Joos 84} following
Pauli, without specifying the bath.

Assuming that the distribution of the field particles is stationary in time,
and homogeneous in space, we can express the trace of the RHS of Eq.(\ref
{3.8}) in terms of the {\em density - density correlation function (CF)} of
the bath,

\begin{eqnarray}
\ Tr_{B}\{\ \rho _{B}\ n({\bf q},t+\tau )\ n({\bf q^{\prime }},t)\}
&=&\langle \ n({\bf q},t+\tau )\ n({\bf q^{\prime }},t)\rangle  \nonumber \\
&=&\delta _{{\bf q^{\prime }},-{\bf q}}\ g_{{\bf q}}(\tau ),  \nonumber \\
g_{{\bf q}}(\tau ) &=&\langle \ n({\bf q},t+\tau )\ n(-{\bf q},t)\rangle .
\label{3.9}
\end{eqnarray}
Eq.~(\ref{3.8}) using (\ref{3.9}) is written as \cite{CCT 92} Eq.~(B.33).

\begin{eqnarray}
\frac{\Delta \overline{\rho }(t)}{\Delta t} &=&-{\frac{1}{{\hbar ^{2}}}}{%
\frac{1}{{V^{2}}}}\sum_{{\bf q}}\ \left| {\phi ({\bf q})}\right|
^{2}\int_{0}^{\infty }\ d\tau {\frac{1}{{\Delta t}}}\int_{t}^{t+\Delta t}\
dt^{\prime }  \nonumber \\
&\ \ &\times \{g_{{\bf q}}(\tau )(A_{{\bf q}}(t^{\prime })\ A_{-{\bf q}%
}(t^{\prime }-\tau )\ {\overline{\rho }(t)}-A_{-{\bf q}}(t^{\prime }-\tau )\ 
{\overline{\rho }(t)}\ A_{{\bf q}}(t^{\prime }))  \nonumber \\
&\ \ &+g_{-{\bf q}}(-\tau )(\ {\overline{\rho }(t)}\ A_{-{\bf q}}(t^{\prime
}-\tau )\ A_{{\bf q}}(t^{\prime })-A_{{\bf q}}(t^{\prime })\ {\overline{\rho 
}(t)}\ A_{-{\bf q}}(t^{\prime }-\tau ))\}.  \label{3.10}
\end{eqnarray}

We further assume that the reservoir is in {\em thermal equilibrium }in
temperature $T$, and thus $\rho _{B}=e^{-\beta H_{B}}/Z,$ where $\beta
=1/k_{B}T$ is the inverse temperature, $k_{B}$ is the Boltzmann factor, and $%
Z$ is the canonical partition function. Introducing the Fourier transform in
time of the CF,

\begin{equation}
g_{{\bf q}}(t)={\frac{1}{{\ 2\pi }}}\int_{-\infty }^{\infty }\ d\omega
e^{-i\omega t}\ g_{{\bf q}}(\omega ),  \label{3.11}
\end{equation}

\noindent we can write the {\em spectral density} of the CF as

\begin{eqnarray}
g_{{\bf q}}(\omega ) &=&\frac{2\pi \hbar }{Z}\sum_{\mu }e^{-\beta E_{\mu
}}\sum_{\nu }\langle \mu |n({\bf q})|\nu \rangle \langle \nu |n(-{\bf q}%
)|\mu \rangle  \nonumber \\
&&\times \delta (E_{\mu }-E_{\nu }+\hbar \omega ),  \label{3.12}
\end{eqnarray}
in terms of the eigenstates of \ $H_{B},$ i.e., \ $H_{B}\ |\mu \rangle
=E_{\mu }\ |\mu \rangle $. First we Notice that since $n(-{\bf q})$ is the 
{\em Hermitian conjugate} of $n({\bf q}),\;g_{{\bf q}}(\omega )$ is a real
function of ${\bf q},\omega .$ \ Then we observe that $g_{-{\bf q}}(-\omega
)=e^{-\beta \hbar \omega }g_{{\bf q}}(\omega ),$ which is a statement of the 
{\em detail balance} induced by the thermal reservoir on the states of the
pointer.

We consider a simple model for the bath, namely a gas of non-interacting
particles in temperature $T$, and average density $n_{0}.$ The mass $m$ of
the bath particles is taken to be much smaller than that of the pointer,
i.e., $m/M\ll 1.$ To calculate the correlation function we first express the
density operator, \ $n({\bf q})=\sum_{{\bf p}}a_{{\bf p}-{\bf q}}^{\dagger
}\ a_{{\bf p}},$ in terms $a_{{\bf p}}^{\dagger }$ \ and \ $a_{{\bf p}},$
the creation and annihilation operators of a ''field'' particle in the
momentum eigenstate $\mid {\bf p}\rangle ,\ $ of energy $\epsilon _{{\bf p}%
}=\hbar ^{2}p^{2}/2m,$ and momentum $\hbar {\bf p}.$ \ Since in the
interaction representation $\ a_{{\bf p}}(t)=a_{{\bf p}}\ \exp {\ (-i\omega
_{{\bf p}}t)},$ where \ $\epsilon _{{\bf p}}=\hbar {\omega _{{\bf p}}}$, and
in thermal equilibrium \ $\langle a_{{\bf p}-{\bf q}}^{\dagger }\ a_{{\bf p}%
}\ a_{{\bf p}^{\prime }+{\bf q}}^{\dagger }\ a_{{\bf p}^{\prime }}\rangle
=\delta _{{\bf p}^{\prime },{\bf p}-{\bf q}}\ f_{{\bf p}-{\bf q}}(1\pm f_{%
{\bf p}}),$ \ where $\pm $ is for either Bose or Fermi particles, and $f_{%
{\bf p}}$ stands for Bose-Einstein of Fermi-Dirac distributions, the
correlation function is then

\begin{equation}
g_{{\bf q}}(\tau )=\sum_{{\bf p}}f_{{\bf p}-{\bf q}}(1\pm f_{{\bf p}})\ e^{{%
-i(\omega _{{\bf p}}-\omega _{{\bf p}-{\bf q}})\tau }},  \label{3.13}
\end{equation}

\noindent and its spectral density is

\begin{equation}
g_{{\bf q}}(\omega )=2\pi \ \sum_{{\bf p}}f_{{\bf p}}(1\pm f_{{\bf p}+{\bf q}%
})\ \delta ((\omega _{{\bf p}+{\bf q}}-\omega _{{\bf p}})-\omega ),
\label{3.14}
\end{equation}

\noindent We now return to Eq.(\ref{3.10}), and express the reduced density
matrix for the pointer in its momentum representation. We denote by \ $\mid 
{\bf k}\rangle ,$ the eigenstate of momentum $\hbar {\bf k,}$ and energy $E_{%
{\bf k}}=\hbar ^{2}k^{2}/2M,$ and find for the matrix elements of the
pointer's operator $A_{{\bf q}}(t)$ in the IR 
\begin{eqnarray}
\langle {\bf k} &\mid &A_{{\bf q}}(t)\mid {\bf k^{\prime }}\rangle =\langle 
{\bf k}\mid e^{iH_{B}t/\hbar }\ e^{i{\bf q}\cdot {\bf R}}\ e^{-iH_{B}t/\hbar
}\mid {\bf k}^{\prime }\rangle  \nonumber \\
&=&e^{i\Omega _{{\bf k},{\bf k^{\prime }}}t}\ \delta _{{\bf k}^{\prime },%
{\bf k}-{\bf q}},  \label{3.15}
\end{eqnarray}
where $\Omega _{{\bf k},{\bf k^{\prime }}}=(E{_{{\bf k}}-}E{_{{\bf k}%
^{\prime }})/\hbar .}$ \ In the {\em Secular Approximation} the equation of
motion for the reduced density matrix, in the {\em Schr\"{o}dinger Picture}
can be written as

\begin{equation}
{\frac{d}{{dt}}}\langle {\bf k}\mid \rho (t)\mid {\bf k^{\prime }}\rangle -{%
\frac{1}{{i\hbar }}}\langle {\bf k}\mid \lbrack H_{p},\rho (t)]\mid {\bf %
k^{\prime }}\rangle =I_{{\bf k},{\bf k}^{\prime }}(t).  \label{3.16}
\end{equation}

\noindent Here the effect of the environment is given by the {\em collision
term}

\begin{eqnarray}
I_{{\bf k},{\bf k}^{\prime }}(t) &=&-{\frac{1}{{\hbar ^{2}}}}{\frac{1}{{V^{2}%
}}}\sum_{{\bf q}}\ {\mid \phi ({\bf q})\mid }^{2}  \nonumber \\
&&\times \{(G_{{\bf q}}(\Omega _{{\bf k},{\bf k}-{\bf q}})+G_{{\bf q}%
}^{*}(\Omega _{{\bf k^{\prime }},{\bf k^{\prime }}-{\bf q}}))\langle {\bf k}%
\mid \rho (t)\mid {\bf k^{\prime }}\rangle  \nonumber \\
&&-(G_{{\bf q}}(\Omega _{{\bf k}+{\bf q},{\bf k}})+G_{{\bf q}}^{*}(\Omega _{%
{\bf k^{\prime }}+{\bf q},{\bf k^{\prime }}}))\langle {\bf k}+{\bf q}\mid
\rho (t)\mid {\bf k^{\prime }}+{\bf q}\rangle \},  \label{3.17}
\end{eqnarray}
where 
\begin{equation}
G_{{\bf q}}(\omega )=\int_{0}^{\infty }\ d\tau e^{i\omega \tau }\ g_{{\bf q}%
}(\tau ),  \label{3.18}
\end{equation}
is the {\em one-sided} Fourier transform of the correlation function. If we
discard the {\em frequency shifts} due to the bath, we can replace \ in the
above equation $\ G_{{\bf q}}(\omega )$ $\ $by $\ $the real function \ $g_{%
{\bf q}}(\omega )/2.$ To further facilitate the{\bf \ }${\bf p-}$summation
in the \ $g_{{\bf q}}(\omega )^{\prime }s,$ \ we make use of the \ $\delta -$
functions, \ and write, for example 
\begin{eqnarray}
G_{{\bf q}}(\Omega _{{\bf k},{\bf k}-{\bf q}}) &=&\ \frac{\pi m}{h^{2}q}%
\frac{1}{V}\sum_{{\bf p}}f_{{\bf p}}(1\pm f_{{\bf p}+{\bf q}})\   \nonumber
\\
&&\times \delta \left\{ {\bf p}\cdot \widehat{{\bf q}}-Q_{+}({\bf k},{\bf q}%
)\right\} ,  \label{3.19}
\end{eqnarray}
where ${\hat{{\bf q}}}={\bf q}/q$, and

\begin{equation}
Q_{\pm }({\bf k},{\bf q})=\frac{m}{M}\widehat{{\bf q}}{\bf \cdot k-}\frac{q}{%
2}(1\pm \frac{m}{M}).  \label{3.20}
\end{equation}

\subsection{\sl Classical Gas Bath}

We now consider the bath as a classical gas, where the thermal distribution
of the particles is {\em Maxwellian}, namely

\begin{equation}
f_{{\bf p}}=n_{0}({\frac{{\alpha }}{{\pi }}})^{3/2}\ e^{-\alpha p^{2}},
\label{3.21}
\end{equation}
with $\alpha ={\hbar }^{2}/2mk_{B}T$. \ In this case, we have \ $f_{{\bf p}%
}\ll 1,$ and if we replace the summation in Eq.(\ref{3.19}) \ by integration
over ${\bf p,}$ i.e. write \ ${\frac{1}{V}}\sum_{{\bf p}}\Rightarrow \int
d^{3}p/{(2\pi )^{3},}$ we find 
\begin{equation}
G_{{\bf q}}(\Omega _{{\bf k},{\bf k}-{\bf q}})={\frac{1}{{\ (2\pi )^{3}}}}%
n_{0}\sqrt{\frac{{\alpha }}{{\pi }}}\frac{\pi m}{q\hbar ^{2}}\ e^{-\alpha {\
Q_{+}^{2}({\bf k},{\bf q})}},  \label{3.22}
\end{equation}
and similarly for the other $G_{{\bf q}}\,^{\prime }$s of Eq.(\ref{3.17}).
The collision term can be then cast into 
\begin{eqnarray}
I_{{\bf k},{\bf k}^{\prime }}(t) &=&-{\frac{1}{V}}\sum_{{\bf q}}\ {\mid \phi
({\bf q})\mid }^{2}{\frac{2\pi }{{\hbar }}}{\frac{1}{{(2\pi )^{3}}}}n_{0}%
\sqrt{\frac{{\alpha }}{{\pi }}}\frac{m}{q\hbar ^{2}}  \nonumber \\
&&\times \{(\ e^{-\alpha {Q_{+}^{2}({\bf k},{\bf q})}}+\ e^{-\alpha {%
Q_{+}^{2}({\bf k}^{\prime },{\bf q})}})\times \langle {\bf k}\mid \rho
(t)\mid {\bf k^{\prime }}\rangle  \nonumber \\
&&\ \ \ -(\ e^{-\alpha {Q_{-}^{2}({\bf k},{\bf q})}}+\ e^{-\alpha {Q_{-}^{2}(%
{\bf k}^{\prime },{\bf q})}})\times \langle {\bf k}+{\bf q}\mid \rho (t)\mid 
{\bf k^{\prime }}+{\bf q}\rangle \}.  \label{3.23}
\end{eqnarray}
This is still an exact result for a classical gas interacting with the
pointer.

We now make use of the two small parameters in the problem, namely, that the
mass ratio is small, and that the momentum transfer is small,

\begin{equation}
\eta =m/M\ll 1{\rm \ and\ }q\ll k.  \label{3.24}
\end{equation}
First we expand $Q^{2}$ in small $\eta $. From Eq.~(\ref{3.20}) we have \ $%
Q_{\pm }^{2}({\bf k},{\bf q})\Rightarrow {\frac{{q^{2}}}{4}}-\eta ({\bf k}%
\cdot {\bf q}\mp {\frac{{q^{2}}}{2}})+O({\eta }^{2}),$ and thus, to first
order in $\eta $ we can write

\begin{equation}
e^{-\alpha Q_{\pm }^{2}({\bf k},{\bf q})}\to e^{-\alpha {\frac{{q^{2}}}{4}}%
}(1+\alpha \eta ({\bf k}\cdot {\bf q}\mp {\frac{{q^{2}}}{2}})).  \label{3.25}
\end{equation}
Then we expand $\left\langle {\bf k+q}\right| \rho (t)\left| {\bf k^{\prime
}+q}\right\rangle $ in power series of ${\bf q},$ to obtain, up to second
order in \ $q/k$

\begin{eqnarray}
\left\langle {\bf k+q}\right| \rho (t)\left| {\bf k^{\prime }+q}%
\right\rangle &=&\left\langle {\bf k}\right| \rho (t)\left| {\bf k^{\prime }}%
\right\rangle +\ {\bf q}\cdot \left( {\frac{\partial }{{\ \partial {\bf k}}}}%
+{\frac{\partial }{{\ \partial {\bf k}^{\prime }}}}\right) \left\langle {\bf %
k}\right| \rho (t)\left| {\bf k^{\prime }}\right\rangle  \nonumber \\
&&+\ {\frac{1}{2}}{\bf q}\cdot \left( {\frac{\partial }{{\ \partial {\bf k}}}%
}+{\frac{\partial }{{\ \partial {\bf k}^{\prime }}}}\right) {\bf q}\cdot
\left( {\frac{\partial }{{\ \partial {\bf k}}}}+{\frac{\partial }{{\
\partial {\bf k}^{\prime }}}}\right) \left\langle {\bf k}\right| \rho
(t)\left| {\bf k^{\prime }}\right\rangle  \label{3.26}
\end{eqnarray}
\mbox{$>$}%
From the structure of this expansion we realize that it is convenient to
express the density matrix in terms of the vectors ${\bf K}=({\bf k}+{{\bf k}%
^{\prime }})/2,$ \ and \ ${\bf p}=\left( {\bf k}-{{\bf k}^{\prime }}\right)
, $ namely write \ $\left\langle {\bf k}\right| \rho (t)\left| {\bf %
k^{\prime }}\right\rangle =\rho ({\bf k},{\bf k}^{\prime },t)=\rho ({\bf K},%
{\bf p},t).$ Substituting these expressions in Eq.(\ref{3.23}), assuming
that \ ${\phi ({\bf q})}$ is isotropic in ${\bf q}$, and integrating over $%
{\bf q}$, the collision term is simply\ 
\begin{eqnarray}
I_{{\bf k},{\bf k}^{\prime }}(t) &=&\gamma {\frac{\partial }{{\partial {\bf K%
}}}}\cdot \left( {\bf K\;}\rho ({\bf K},{\bf p},t)\right)  \nonumber \\
&&+D{\frac{\partial }{{\partial {\bf K}}}}\cdot {\frac{\partial }{{\partial 
{\bf K}}}\;}\rho ({\bf K},{\bf p},t).  \label{3.27}
\end{eqnarray}
Here 
\begin{equation}
\gamma =n_{0}\eta {\frac{1}{{\ (2\pi )^{4}}}}\ \sqrt{\frac{{\alpha }}{{\pi }}%
}\ \frac{4m\alpha }{3{{\hbar }^{3}}}\int_{0}^{\infty }\ dq\ q^{3}\ \left| {%
\phi (}q)\right| ^{2}\ e^{-\alpha {\frac{{q^{2}}}{4}}},  \label{3.28}
\end{equation}
plays the role of {\em inverse relaxation time}, and $D=\gamma /2\alpha \eta 
$ plays the role of a {\em diffusion coefficient in k space}. Using \cite
{Chandrasekhar 43} we can relate the $k-$diffusion coefficient, $D,$ in Eq. (%
\ref{3.27}) to the standard coefficient of spatial diffusion, $%
D_{c}=k_{B}T/M\gamma ,$ i.e., we have $D=(M\gamma /\hbar )^{2}D_{c}.$ Notice
that it is spatial diffusion coefficient, $D_{c},$ which is connected by the
Einstein relation to $\gamma .$

The equation of motion of density matrix of the pointer, Eq.~(\ref{3.16}),
becomes, with Eq.~(\ref{3.27}), a {\em Fokker - Planck (FP) equation} for $%
\langle {\bf k}\mid \rho (t)\mid {\bf k^{\prime }}\rangle ,$ namely

\begin{eqnarray}
\frac{\partial }{\partial t}\left\langle {\bf k}\right| \rho (t)\left| {\bf k%
}^{\prime }\right\rangle -\frac{1}{i\hbar }\left\langle {\bf k}\right|
[H_{P},\rho (t)]\left| {\bf k}^{\prime }\right\rangle &=&\gamma \frac{%
\partial }{\partial {\bf K}}\cdot \left( {\bf K\;}\left\langle {\bf k}%
\right| \rho (t)\left| {\bf k}^{\prime }\right\rangle \right)  \nonumber \\
&&+D\frac{\partial }{\partial {\bf K}}\cdot \frac{\partial }{\partial {\bf K}%
}\;\left\langle {\bf k}\right| \rho (t)\left| {\bf k}^{\prime }\right\rangle
.  \label{3.29}
\end{eqnarray}
This equation is much like the FP equation derived by Caldeira and Lggett 
\cite{Caldeira 83} for a harmonic oscillator interacting with a bath of
harmonic oscillators in equilibrium. The main difference is that the
friction parameter $\gamma ,$ here is given in terms of the properties of
the gas surrounding the pointer.

Before we turn to the solution of this equation in time, we make some
estimates of the Fokker-Planck coefficient, $\gamma .$ If the interaction
potential ${\phi (}{\bf r})$ is specified, we can calculate explicitly $%
\gamma \ $of Eq.(\ref{3.28}).\ Let us propose a particular interaction
potential that will simulate the collisions between the pointer and the gas
particles. We take

\begin{equation}
\phi ({\bf r})=\phi _{0}e^{-r^{2}/a^{2}},  \label{3.29.1}
\end{equation}
where $a$ is the length scale of the pointer, and $\phi _{0}$ is the
potential strength. Using this potential, the term $\left| {\phi (}q{)}%
\right| ^{2}$ in Eq. (\ref{3.28}) can be specified as

\begin{equation}
\left| {\phi (}q{)}\right| ^{2}=\left| \int \phi ({\bf r})e^{-i{\bf q\cdot r}%
}d{\bf r}\right| ^{2}=\pi ^{3}a^{6}\phi _{0}^{2}e^{-a^{2}q^{2}/2}.
\label{3.29.6}
\end{equation}
For the inverse{\it \ }relaxation time we get

\begin{equation}
\gamma =n_{0}\eta \sqrt{\frac{{\alpha }}{{\pi }^{3}}}\ \frac{m}{3{{\hbar }%
^{3}}}\left( \phi _{0}a^{2}\right) ^{2}\frac{\varrho }{\left( 1+\varrho
\right) ^{2}},  \label{3.29.7}
\end{equation}
where $\varrho =2a^{2}/{\alpha .}$ Recalling that $\alpha ={\hbar }%
^{2}/2mk_{B}T,$ and noting that even for objects with atomic dimensions $%
\varrho \gg 1,$ we can write $\gamma $ as

\begin{equation}
\gamma =\frac{1}{16}\sqrt{\frac{3}{\ 2\pi ^{3}}}n_{0}\eta a^{2}\overline{v}%
\left( \frac{\phi _{0}}{\epsilon _{T}}\right) ^{2},  \label{3.29.8}
\end{equation}
where $\epsilon _{T}=3k_{B}T/2$ is the average thermal energy and $\overline{%
v}=\sqrt{3k_{B}T/m}$ is the average velocity of the gas particles. Although
we have an explicit expression for $\gamma $, it is still necessary to
specify $\phi _{0}.$ For air in room temperature, a pointer which is a
silver atom, we take $a=1.75\times 10^{-8}$ ${\rm cm}$ to be the radius of
the atom, and $M=1.8\times 10^{-22}$ ${\rm g}$. Assuming that the barrier $%
\phi _{0}$ is, say 50 times the thermal energy $\epsilon _{T}$, we get a
typical $\gamma \simeq 2.5\times 10^{9}{\rm s}^{{\rm -1}}.$

\subsection{\sl Solution of the Fokker-Planck Equation for the Pointer in a
Bath.}

In order to solve Eq. (\ref{3.29}) for the pointer model outlined in section
II, we further make two simplifying assumptions. First, we observe that
since the deflection of the pointer as a result of the interaction with the
spin is only along the $x$-axis, the coordinates $y$ and $z$ are of little
consequence for our purpose. Hence we will consider only the one dimensional
FP equation along the{\em \ x-coordinate, }and use, for convenience, $x$
instead of the $X-$coordinate of the pointer. If we set ${\bf k}\rightarrow
k,$ ${\bf k}^{\prime }\rightarrow k^{\prime },\;K=(k+k^{\prime })/2,$ and \ $%
p=(k-k^{\prime })$\ we can restrict ourselves only to

\begin{equation}
\left\langle k\right| \rho (t)\left| k^{\prime }\right\rangle =\rho
(k,k^{\prime },t)=\rho (K,p,t).  \label{3.30}
\end{equation}
Our second assumption is that the interaction of the pointer with the atom
occurs in a very short time, or takes place before the interaction with the
bath is switched on. In other words, we start from a density matrix which
describes a {\em free} pointer {\em after} its interaction with the spin has
already been completed, that is we use Eq. (\ref{2.9}) as the initial
condition for the FP equation for $\rho (K,p,t)$. It is clear that both
assumptions are not necessary in order to carry out the calculation. The
first one comes to simplify the mathematics by removing the irrelevant
degrees of freedom, while the second one saves us the need to define $V(t)$
explicitly in Eq. (\ref{1.3}). We bare in mind of course, that what we are
really interested in here, is the decoherence process itself, and not so
much the creation of the initial coherent state of the pointer.

Suppose therefore that we know the initial density matrix, $\rho (p,K,0),$
and we seek the solution of $\rho (p,K,t)$ at the time $t$. The {\em free
streaming} term of Eq.~(\ref{3.29}) for the pointer in its momentum
representation is then

\begin{eqnarray}
-\frac{1}{i\hbar }\left\langle k\right| [H_{P},\rho (t)]\left| k^{\prime
}\right\rangle &=&-\frac{1}{i\hbar }(E_{k}-E_{k^{\prime }})\rho (k,k^{\prime
},t)  \nonumber \\
&=&i\hbar \frac{{\ }Kp}{M}\ \rho (p,K,t),  \label{3.31}
\end{eqnarray}
Inserting this back into Eq.~(\ref{3.29}) gives us

\begin{eqnarray}
{\frac{\partial }{{\partial t}}}\rho (p,K,t)+i\hbar \frac{{\ }Kp}{M}\ \rho
(p,K,t) &=&\gamma \ {\frac{\partial }{{\ \partial }K}}\ \left( K\ \rho
(p,K,t)\right)  \nonumber \\
&&+D{\frac{{{\partial }^{2}}}{{\ \partial K^{2}}}}\rho (p,K,t).  \label{3.32}
\end{eqnarray}
where the initial condition can be written with the aid of (\ref{2.9}) and (%
\ref{2.10}) as

\begin{eqnarray}
\rho (p,K;\sigma ,{\sigma }^{\prime },0) &=&a_{{\sigma }^{\prime
}}^{*}a_{\sigma }2(2\pi {\Delta }^{2})^{1/2}e^{-2{\Delta }^{2}K^{2}-iK%
\overline{X}(\sigma -{\sigma }^{\prime })}  \nonumber \\
&&\times e^{-{\Delta }^{2}p^{2}/2-ip\overline{X}(\sigma +{\sigma }^{\prime
})/2}.  \label{3.33}
\end{eqnarray}
Since the measurement is done after the interaction was completed, namely at
a time $t>T$, we set $\overline{X}(t)=\overline{X}=\int_{0}^{T}\ dt^{\prime
}V(t^{\prime }).$

The method of solving Eq. (\ref{3.32}) with the initial condition (\ref{3.33}%
) is outlined in appendix B. Here we just quote the result, which is

\begin{eqnarray}
\rho (p,K;\sigma ,{\sigma }^{\prime };t) &=&a_{{\sigma }^{\prime }}^{\ast
}a_{\sigma }e^{\gamma t}\ \sqrt{\frac{8\pi u{\Delta }^{2}}{2{\Delta }^{2}+u}}%
e^{-{\Delta }^{2}p^{2}/2-ip\overline{X}(\sigma +{\sigma }^{\prime
})/2-D\Theta p^{2}}e^{-i\frac{\hbar }{2M}\lambda Kp-ue^{2\gamma t}K^{2}} 
\nonumber \\
&&\times \exp \left\{ -\frac{\left[ \overline{X}(\sigma -{\sigma }^{\prime
})+\frac{\hbar }{2M}\lambda p+i2ue^{\gamma t}K\right] ^{2}}{4\left( 2{\Delta 
}^{2}+u\right) }\right\} ,  \label{3.34}
\end{eqnarray}
where we have defined

\begin{eqnarray}
u(t) &\equiv &1/4D\eta (t),  \nonumber \\
\lambda (t) &\equiv &\frac{\zeta ^{2}(t)}{\eta (t)}=\frac{{2}}{{\ \gamma }}%
\frac{\left( {e^{\gamma t}-1}\right) }{\left( {e^{\gamma t}+1}\right) }, 
\nonumber \\
\Theta (t) &\equiv &\left( \frac{\hbar }{M\gamma }\right) ^{2}\left[
t-\lambda (t)\right] .  \label{3.35}
\end{eqnarray}

Once we have this solution we return to the outcome of a {\em measurement}
of the position of the pointer. In the present one-dimensional case the {\em %
probability density} to detect the center of mass of the pointer around $x,$
is determined, as in the previous ''free'' pointer case by\ $%
P(x,t)=\sum_{\sigma ,\sigma ^{\prime }}\rho (x,x,\sigma ,\sigma ^{\prime
};t),$ namely by the diagonal elements of $\rho $ in real space. We thus
have to transform Eq. (\ref{3.34}) back to $(x,x^{\prime })$ space and then
to set \ $x=x^{\prime }$ in the resulting density matrix. This can be done
conveniently by noting that the Fourier transform from $(k,k^{\prime })$
space to $(x,x^{\prime })$ space i.e.

\begin{equation}
\rho (x,x^{\prime },t)=\int \frac{dk}{2\pi }e^{-ik^{\prime }x^{\prime }}\int 
\frac{dk}{2\pi }e^{ikx}\rho (k,k^{\prime },t),  \label{3.36}
\end{equation}
can be written with new variables, $\theta =(x-x^{\prime })$ and $%
X=(x+x^{\prime })/2,$ and the use of the definitions $K=(k+k^{\prime })/2$
and $p=(k-k^{\prime }),$ as

\begin{equation}
\rho (X,\theta ,t)=\int \frac{dp}{2\pi }e^{ipX}\int \frac{dK}{2\pi }%
e^{iK\theta }\rho (K,p,t).  \label{3.37}
\end{equation}
Now the probability density will be given simply by $P(x,t)=\sum_{\sigma
,\sigma ^{\prime }}\rho (X=x,\sigma ;\theta =0,\sigma ^{\prime };t).$ We
write $P(x,t)$ as a sum of three terms:

\begin{equation}
P(x,t)=P_{+1}(x,t)+P_{-1}(x,t)+P_{+1,-1}(x,t).  \label{3.38}
\end{equation}
where\ $P_{\pm 1}(x,t)=$ $\rho (x,x,\pm 1,\pm 1;t)$\ stand for the diagonal
elements, and\ $P_{+1,-1}(x,t)=\rho (x,x,+1,-1;t)+\rho (x,x,-1,+1;t)$\
stands for the off-diagonal elements.

The first and second terms of \ Eq. (\ref{3.38}) are the probabilities of
the spin-up and spin-down components of the measured atom. Performing the
transformation in Eq. (\ref{3.37}) we get for $\sigma ^{\prime }=\sigma =+1$%
\begin{equation}
P_{+1}(x,t)=\left| a_{+1}\right| ^{2}\sqrt{\frac{1}{2\pi \Delta _{\beta
}^{2}(t)}}\ \exp \left\{ -\frac{(x-\overline{X})^{2}}{2\Delta _{\beta
}^{2}(t)}\right\} ,  \label{3.39a}
\end{equation}
and for \ $\sigma ^{\prime }=\sigma =-1$%
\begin{equation}
P_{-1}(x,t)=\left| a_{-1}\right| ^{2}\sqrt{\frac{1}{2\pi \Delta _{\beta
}^{2}(t)}}\ \exp \left\{ -\frac{(x+\overline{X})^{2}}{2\Delta _{\beta
}^{2}(t)}\right\} .  \label{3.39b}
\end{equation}
The spatial broadening of the pointer's position, which is due to both the
free quantum diffusion and the effect of the bath, is expressed as 
\begin{equation}
\Delta _{\beta }^{2}(t)=\Delta ^{2}[\varkappa (t)+\kappa (t)],  \label{3.40}
\end{equation}
where 
\begin{equation}
\varkappa (t)=1+\left( \frac{1}{{\gamma }\tau _{f}}\right) ^{2}(1-e^{-\gamma
t})^{2},  \label{3.40.1}
\end{equation}
and 
\begin{equation}
\kappa (t)=\left( \frac{1}{\gamma \tau _{f}}\right) ^{2}\frac{4{\Delta }^{2}D%
}{{\gamma }}f(\gamma t)  \label{3.40.2}
\end{equation}
is expressed in terms of the function 
\begin{equation}
f(x)=2\left( x-2\frac{(1-e^{-x})}{\left( 1+e^{-x}\right) }\right) +\frac{%
\left( 1-e^{-x}\right) ^{3}}{{\ }\left( 1+e^{-x}\right) }.  \label{3.40.3}
\end{equation}
We now compare these expressions with those obtained previously, Eqs. (\ref
{2.18a}), (\ref{2.18b}) and\ (\ref{2.20}), in the absence the bath. We
observe that the previous broadening of the pointer's position,\ $\Delta
_{f}(t),$ is replaced by $\Delta _{\beta }(t),$ where $\varkappa (t)$ takes
care of both the free quantum diffusion and the {\em friction} caused by the
environment, and $\kappa (t)$ corresponds to the {\em diffusion} afflicted
on the pointer by the gas surrounding it. The effect of the environment on
the probabilities $P_{\pm 1}(x,t)$\ is just to enlarge the uncertainty of
the pointer's position.

Now let us look at the off-diagonal term of \ Eq. (\ref{3.38}). Taking $%
a_{\pm 1}=\left| a_{\pm 1}\right| e^{i\varphi _{\pm }}$ we get

\begin{eqnarray}
P_{+1,-1}(x,t) &=&2\sqrt{P_{+1}(x,t)P_{-1}(x,t)}\exp \left\{ -\frac{%
\overline{X}^{2}\kappa (t)}{2\Delta _{\beta }^{2}(t)}\right\}  \nonumber \\
&&\times \cos \left( \frac{x\,\overline{X\,}}{\Delta _{\beta }^{2}(t)\,\tau
_{f}}\frac{1}{\gamma }\,(1-{e^{-\gamma t}})+\varphi _{-}-\varphi _{+}\right)
,  \label{3.42}
\end{eqnarray}
where\ $\kappa (t)$ is given by Eq.(\ref{3.40.2}). This is the main result
of the present paper. The function $P_{+1,-1}(x,t)$ \ represents the
probability of finding the system in a state of interference between the two
possible ''classical'' outcomes of the measurement. If we compare this
result with the interference term, Eq. (\ref{2.19}), \ of the ''free''
pointer, we observe two changes: (i) the argument of the $\cos $ in \ Eq.(%
\ref{2.19}) is replaced by a more elaborate time dependent expression in Eq.(%
\ref{3.42}), yet in the limit $\gamma \rightarrow 0$ it reduces to the
expression in Eq.(\ref{2.19}), and (ii) a new time dependent exponential
factor appears in\ Eq.(\ref{3.42}). This exponential factor is the important
effect of the environment on the pointer, and we examine it now more
closely. Let us write it as $e^{-g(t)},$ where $\ g(t)$ \ is the {\em %
decoherence function}, and it is given by

\begin{equation}
g(t)=\frac{X_{0}^{2}\kappa (t)}{2\Delta _{\beta }^{2}(t)}=\frac{\overline{X}%
^{2}}{2{\Delta }^{2}}\left[ \frac{\kappa (t)}{\kappa (t)+\ \varkappa (t)}%
\right] .  \label{3.44}
\end{equation}
Notice that it is not the friction, which is due to energy exchange with the
bath, that is responsible for the decoherence, but rather the diffusion in
momentum space, which is originating it.

For times much smaller than the relaxation time, i.e. $\gamma t<<1,$ we have

\begin{equation}
\kappa (t)\rightarrow D\left( \frac{1}{\gamma \tau _{f}}\right) ^{2}\frac{{%
\Delta }^{2}}{{\gamma }}\frac{8}{3}({\gamma }t)^{3}=2D\left( \frac{2{\Delta }%
}{\tau _{f}}\right) ^{2}\frac{t^{3}}{3},  \label{3.47}
\end{equation}
and since $\varkappa (t)$ is of order 1,

\begin{equation}
g(t)\rightarrow \frac{\overline{X}^{2}}{{\Delta }^{2}}D\left( \frac{2{\Delta 
}}{\tau _{f}}\right) ^{2}\frac{t^{3}}{3}.  \label{3.48}
\end{equation}
Remembering that $D=M\gamma k_{B}T/\hbar ^{2}$, we write $g(t)$ $\ $as $\
(\Gamma ^{\prime }t)^{3},$ where

\begin{equation}
\Gamma ^{\prime }=\left( \frac{\overline{X}^{2}}{{\Delta }^{4}}\frac{\gamma
k_{B}T}{3M}\right) ^{1/3}=\left[ \frac{1}{3}\frac{\overline{X}^{2}}{{\Delta }%
^{2}}\frac{\gamma k_{B}T}{\hbar ^{2}}\left( \frac{\hbar ^{2}}{M{\Delta }^{2}}%
\right) \right] ^{1/3}  \label{3.50}
\end{equation}
plays the role of a kind of inverse decay time of the interference. It is
instructive to compare $\Gamma ^{\prime }$\ of Eq. (\ref{3.50}) with the
analogous expression of Ref. \cite{Caldeira 85}, Eq.(3.5). Their result was
obtained for an oscillator, with a typical frequency of $\omega _{R},$ which
is coupled to a reservoir of harmonic oscillators, in the high temperature
limit, for the weakly damped case, \ $\gamma /\omega _{R}\ll 1,$ and at
initial times. The only difference between their $\Gamma _{2},$ and that of
our system, $\Gamma ^{\prime },$ is that their oscillator's quanta $\hbar
\omega _{R}$ is replaced by the energy term $\hbar ^{2}/M{\Delta }^{2},$
which is related to the quantum mechanical energy associated with the
initial spread of the pointer. To make a simple estimate for $\Gamma
^{\prime }$ let us take our pointer as a silver atom. We thus take ${\Delta
\sim 1}$ ${\rm \mu m}$, to represent a pinhole for localizing the atom, $%
\overline{X}\sim 1$ {\rm cm,} as a deflection position, $\ M\sim 1.8\times
10^{-22}$ ${\rm g}$,{\rm \ }and $a=1.75\times 10^{-8}$ ${\rm cm.}$ For air
at room temperature, namely with $\tau _{r}=1/\gamma \simeq 0.4${\rm \ n}$%
{\rm s,}$ we obtain $\Gamma ^{\prime }\tau _{r}\sim 50.$ We see that the
typical decoherence time $1/\Gamma ^{\prime }$ is almost two orders of
magnitude smaller then relaxation time. However this behavior does not
proceed for times much longer than the relaxation time. When time increases,
so does $g(t)$, and eventually it reaches a{\em \ saturation} value of $%
g(t)\rightarrow {\frac{1}{2}}\left( \overline{X}/\Delta \right) ^{2},$ which
is of the order of $10^{8}.$ Nevertheless, at later times when $\gamma t>1,$
but long before saturation, we have

\begin{equation}
\kappa (t)\rightarrow 2D\left( \frac{2\Delta }{\gamma \tau _{f}}\right)
^{2}t,  \label{3.51}
\end{equation}
which gives

\begin{equation}
g(t)\rightarrow \frac{\overline{X}^{2}}{2{\Delta }^{2}}\left[ \frac{2D\left( 
\frac{2{\Delta }}{\gamma \tau _{f}}\right) ^{2}t}{2D\left( \frac{2{\Delta }}{%
\gamma \tau _{f}}\right) ^{2}t+{1}}\right] .  \label{3.52}
\end{equation}
At this time range, $g(t)$ is approximately linear in time, i.e.,\ $%
g(t)\rightarrow \Gamma t,$\ with a slope given by

\begin{equation}
\Gamma =\frac{\overline{X}^{2}}{{\Delta }^{4}}\frac{k_{B}T}{M\gamma }=\frac{%
\overline{X}^{2}}{{\Delta }^{2}}\frac{k_{B}T}{\hbar ^{2}\gamma }\frac{\hbar
^{2}}{M{\Delta }^{2}},  \label{3.53}
\end{equation}
for the silver atom it yields $\Gamma /\gamma =3.5\times 10^{5}.$ Thus the
typical decoherence time $1/\Gamma $ in the linear part is much shorter then 
$1/\Gamma ^{\prime }.$ The meaning of this result, as was pointed out by
Ref. \cite{Caldeira 85}, is that as far as the decoherence is concerned, the
system becomes aware of the bath, in a time scale much shorter than $\tau
_{r},$ and the interference is deteriorating almost immediately. Our
decoherence rate,\ $\Gamma ,$ of \ Eq. (\ref{3.53}), for the gas bath, is
quite similar to that of the oscillators' bath of Ref. \cite{Caldeira 85},
Eq.(3.7), at the high temperature and strongly damped limit. It is
interesting to note that the strongly damped limit, which is defined as $%
R_{osc}=\gamma /\omega _{R}\gg 1$ for the oscillators' bath, is replaced
here by $R_{f}=\gamma \tau _{f}\gg 1$. In our analysis the meaning of \ the
''strongly damped'' case is not clear, yet taking $\hbar ^{2}/M{\Delta }^{2}$
instead of the oscillator's quanta $\hbar \omega _{R}$ of Ref. \cite
{Caldeira 85}, we see that $R_{f}=\gamma \tau _{f}\sim 10^{7}$ is indeed
very large. So in that sense we are in the strongly damped limit. It is also
interesting to compare $\Gamma $ \ of \ Eq. (\ref{3.53}) \ with the
decoherence rate, \ $\Gamma _{Z}\simeq \gamma (\overline{X}/\lambda
_{T})^{2},$ quoted by Zurek Ref. \cite{Zurek 91}, where \ $\lambda
_{T}=\hbar /\sqrt{2Mk_{B}T\text{ }}$ \ \ is the thermal de Broglie wave
length of the pointer. Apparently Zurek assumes that the decoherence time,
is proportional to the inverse of the coefficient of the last term in his
equation (9). We find that the ratio $\Gamma /\Gamma _{Z}$ \ is of order \ $%
1/R_{f}^{2}$, i.e., the decoherence time due to the interaction of the
pointer with the gas is much longer than Zurek estimated. This is also the
case for the Caldeira and Leggett \cite{Caldeira 85} decoherence rates, $%
\Gamma _{1}$ \ and \ $\Gamma _{3}$: The ratio $\Gamma _{1}/\Gamma _{Z}$ \ in
the weakly damped limit, is of order \ $R_{ocs}/R_{f},$ while in the
strongly damped limit, \ $\Gamma _{3}/\Gamma _{Z}$ \ is of order \ $%
1/R_{f}R_{osc}.$

Fig. 1 shows a graphical description summarizing the discussion above. The
decoherence function, $g(t)$ of Eq. (\ref{3.44}), normalized to the
saturation value ${\frac{1}{2}}\left( \overline{X}/\Delta \right) ^{2}$ is
depicted as a function of $\gamma t.$ In the inset, we see more closely the
behavior at initial times, where $g(t)$ starts as $(\Gamma ^{\prime }t)^{3}$
and quickly goes into the linear behavior as $\Gamma t.$

\section{Conclusions}

In this paper we have studied the affect of the environment on the process
of measurement of the state of a microscopic spin particle by a classical
pointer. It has been shown that, the entangled state, which occurred due to
the interaction of the spin particle with the measuring apparatus, the
pointer, is developed into a mixed state by the environment, in an extremely
short time. An initially prepared state of a spin half particle, as a
superposition of up and down states, is transformed by the interaction into
corresponding two distinguishable positions of the pointer. The probability
to find the pointer in these positions is expressed in terms of two
separated wave-packets, and an interference term, which is a manifestation
of the entanglement between the particle and the pointer. Due to the
coupling of the pointer with the environment, which is modeled here as a gas
of independent particles, rather than an ensemble of harmonic oscillators,
this quantum interference decays extremely fast in time. The final mixed
state, which is the outcome of the decoherence, introduced by the
environment, is being expressed as the two separated wave-packets of the
pointer's position. The probabilistic nature of quantum mechanics is still
maintained in the final representation of the pointer, but the interference
is destroyed. An exponential decay function, which portrays the decoherence,
is expressed in terms of the parameters of the pointer and the bath.

\section{Appendix A}

\subsection{\it Solving the Fokker-Planck Equation for the Pointer}

In the present Appendix we seek the solution for Eq. (\ref{3.32}) with the
initial condition (\ref{3.33}). It will be convenient to study the density
matrix as a {\em Wigner distribution function}, which is defined by the
transformation

\begin{equation}
\rho (K,p,t)=\int_{-\infty }^{\infty }\ dX\ e^{-iXp}\ W(X,K,t).  \label{A.1}
\end{equation}
Eq.~(\ref{3.32}) is transformed to the equation

\begin{eqnarray}
{\frac{\partial }{{\partial t}}}W(X,K,t) &+\frac{\hbar K}{M}&\frac{\partial 
}{\partial X}\ W(X,K,t)  \nonumber \\
&=&\gamma \ \frac{\partial }{{\ \partial }K}\ \left( K\ W(X,K,t)\right) +D%
\frac{{{\partial }^{2}}}{{\ }\partial K^{2}}\ W(X,K,t),  \label{A.2}
\end{eqnarray}
where $W(X,K,t)$ \ plays a role analogous to that of a classical
distribution function, $f(x,p,t)$ \ to find a massive test particle near the
position \ $x$ \ and the momentum $p,$ while moving in a gaseous bath. This
partial differential equation can be solved by the use of the method of
characteristics. We start by ignoring the term with the second order
derivative in $K$, and we look for curves that satisfy the conditions

\begin{equation}
\frac{dX}{\hbar K/M}=dt=-\frac{dK}{\gamma K}=\frac{dW}{\gamma W}.
\label{A.3}
\end{equation}
Integrating with respect to $t$ we have

\begin{equation}
dt=-\frac{dK}{\gamma K}\rightarrow K=\kappa e^{-\gamma t}\rightarrow \kappa
=Ke^{\gamma t}  \label{A.4}
\end{equation}

\begin{equation}
\frac{dX}{\hbar K/M}=dt\rightarrow X=-\frac{\hbar }{\gamma M}\kappa
e^{-\gamma t}+\xi \rightarrow \xi =X+\frac{\hbar K}{M\gamma }  \label{A.5}
\end{equation}

\begin{equation}
dt=\frac{dW}{\gamma W}\rightarrow W=\omega e^{\gamma t}  \label{A.6}
\end{equation}
where we wrote the constants of integration over $t,$ i.e. $\kappa ,\xi $
and $\omega ,$ in terms of the variables $K,X,t.$

We can now make the independent variables' transformation $K,X,t\to \kappa
,\xi ,\tau $, where

\begin{eqnarray}
\kappa &=&Ke^{\gamma t},  \nonumber \\
\xi &=&X+\frac{\hbar K}{M\gamma }  \nonumber \\
\tau &=&t  \label{A.7}
\end{eqnarray}
and find that in terms of the new variables, and the substitution 
\begin{equation}
W(\xi ,\kappa ,\tau )=\omega (\xi ,\kappa ,\tau )\ e^{\gamma \tau },
\label{A.10}
\end{equation}
Eq.~(\ref{A.2}) takes the form 
\begin{equation}
{\frac{\partial }{{\partial \tau }}}\omega (\xi ,\kappa ,\tau )=D\left(
e^{\gamma \tau }{\frac{\partial }{{\ \partial \kappa }}}+{\frac{\hbar }{{\
M\gamma }}}{\frac{\partial }{{\ \partial \xi }}}\right) ^{2}\omega (\xi
,\kappa ,\tau ).  \label{A.11}
\end{equation}
This equation is a diffusion equation in $\xi ,\kappa ,\tau $ space, with a
time dependent diffusion coefficient. This can be solved for example by
using Fourier transform in $\xi ,\kappa $ space, namely, introducing $%
f(p,x,\tau )$ by

\begin{equation}
f(p,x,\tau )={\frac{1}{{2\pi }}}\int_{-\infty }^{\infty }\ d\xi \ \ e^{-i\xi
p}\int_{-\infty }^{\infty }\ dx\ e^{i\kappa x}\omega (\xi ,\kappa ,\tau )
\label{A.12}
\end{equation}
Eq.~(\ref{A.11}) is simply replaced by

\begin{equation}
{\frac{\partial }{{\partial \tau }}}f(p,x,\tau )=-D\left( e^{\gamma {\tau }%
}x-{\frac{\hbar }{{\ M\gamma }}}p\right) ^{2}\ f(p,x,\tau ).  \label{A.13}
\end{equation}
with the formal solution

\begin{equation}
f(p,x,\tau )=f(p,x,0)\exp \left\{ -D\int_{0}^{\tau }d{\tau }^{\prime }\left(
e^{\gamma {\tau }^{\prime }}x-{\frac{\hbar }{{\ M\gamma }}}p\right)
^{2}\right\} ,  \label{A.14}
\end{equation}
where $f(p,x,0)$ is the initial value of $f.$ If we denote by

\begin{eqnarray}
\eta (t) &=&\int_{0}^{t}\ dt^{\prime }e^{2\gamma t^{\prime }}={\frac{{\
e^{2\gamma t}-1}}{{\ 2\gamma }}}  \nonumber \\
\zeta (t) &=&\int_{0}^{t}\ dt^{\prime }e^{\gamma t^{\prime }}={\frac{{\
e^{\gamma t}-1}}{{\ \gamma }}},  \label{A.15}
\end{eqnarray}
we can write Eq.~(\ref{A.14}) as

\begin{equation}
f(p,x,\tau )=f(p,x,0)\exp \left\{ -D\left( \eta (\tau )x^{2}-2\zeta (\tau ){%
\frac{\hbar }{{\ M\gamma }}}xp+({\frac{\hbar }{{\ M\gamma }}})^{2}\tau
p^{2}\right) \right\} .  \label{A.16}
\end{equation}

We return now to Eq.~(\ref{A.12}) to find

\begin{eqnarray}
\omega (\xi ,\kappa ,\tau ) &=&{\frac{1}{{\ 2\pi }}}\int_{-\infty }^{\infty
}\ dp\ e^{ip\xi }\int_{-\infty }^{\infty }\ dx\ e^{-ix\kappa }  \nonumber \\
&&e^{-D\left( \eta (\tau )x^{2}-2\zeta (\tau ){\frac{\hbar }{{\ M\gamma }}}%
xp+({\frac{\hbar }{{\ M\gamma }}})^{2}\tau p^{2}\right) }\ f(p,x,0) 
\nonumber \\
&=&{\frac{1}{{\ (2\pi )^{2}}}}\int_{-\infty }^{\infty }\ dp\ e^{ip\xi
}\int_{-\infty }^{\infty }\ dx\ e^{-ix\kappa }\int_{-\infty }^{\infty }\ d{%
\xi }^{\prime }\ e^{-i{\xi }^{\prime }p}\int_{-\infty }^{\infty }\ d{\kappa }%
^{\prime }\ e^{i{\kappa }^{\prime }x}  \nonumber \\
&&e^{-D\left( \eta (\tau )x^{2}-2\zeta (\tau ){\frac{\hbar }{{\ M\gamma }}}%
xp+({\frac{\hbar }{{\ M\gamma }}})^{2}\tau p^{2}\right) }\ \omega ({\xi }%
^{\prime },{\kappa }^{\prime },0).  \label{A.17}
\end{eqnarray}
Note that at $t=0$ we have $W(\xi ,\kappa ,0)=\omega (\xi ,\kappa ,0)$ and
the transformation in Eq. (\ref{A.7}) has the form

\begin{eqnarray*}
\kappa ^{\prime } &=&K^{\prime } \\
\xi ^{\prime } &=&X^{\prime }+\frac{\hbar K^{\prime }}{M\gamma },
\end{eqnarray*}
hence we can write Eq. (\ref{A.17}) in terms of the Wigner distribution

\begin{eqnarray}
W(\xi ,\kappa ,\tau ) &=&e^{\gamma \tau \ -i\kappa (\sigma -\sigma ^{\prime
})X_{0}}{\frac{1}{{\ 2\pi }}}\int_{-\infty }^{\infty }\ dp\ e^{ip\xi
}\int_{-\infty }^{\infty }\ dx\ e^{-ix\kappa }  \nonumber \\
&&e^{-D\left( \eta (\tau )x^{2}-2\zeta (\tau ){\frac{\hbar }{{\ M\gamma }}}%
xp+({\frac{\hbar }{{\ M\gamma }}})^{2}\tau p^{2}+(\sigma -\sigma ^{\prime
})X_{0}\left[ \eta (\tau )(\sigma -\sigma ^{\prime })X_{0}-2\zeta (\tau )%
\frac{\hbar }{M\gamma }p+2\eta (\tau )x\right] \right) }\ f(p,x,0)  \nonumber
\\
&=&{\frac{1}{{\ (2\pi )^{2}}}}e^{\gamma t\ -iKe^{\gamma t}(\sigma -\sigma
^{\prime })X_{0}}\int_{-\infty }^{\infty }\ dp\ e^{ip\left( X+\frac{\hbar K}{%
M\gamma }\right) }\int_{-\infty }^{\infty }\ dx\   \nonumber \\
&&\times e^{-ixKe^{\gamma t}}e^{-D\left( \eta (t)x^{2}-2\zeta (t){\frac{%
\hbar }{{\ M\gamma }}}xp+({\frac{\hbar }{{\ M\gamma }}})^{2}tp^{2}+(\sigma
-\sigma ^{\prime })X_{0}\left[ \eta (t)(\sigma -\sigma ^{\prime
})X_{0}-2\zeta (t)\frac{\hbar }{M\gamma }p+2\eta (t)x\right] \right) } 
\nonumber \\
&&\int_{-\infty }^{\infty }\ dX^{\prime }\ e^{-i\left( X^{\prime }+\frac{%
\hbar K^{\prime }}{M\gamma }\right) p}\int_{-\infty }^{\infty }\ dK^{\prime
}e^{iK^{\prime }x}\ W(X^{\prime },K^{\prime },0).  \label{A.18}
\end{eqnarray}

\noindent Using Eq.~(\ref{A.7}) and Eq.~(\ref{A.10}) the Wigner distribution
at time $t$ can be then expressed in terms of its initial value as

\begin{equation}
W(X,K,t)={\frac{1}{{2\pi }}}\int_{-\infty }^{\infty }\ dX^{\prime }\
\int_{-\infty }^{\infty }\ dK^{\prime }\ J(X,K,t;X^{\prime },K^{\prime },0)\
W(X^{\prime },K^{\prime },0),  \label{A.19}
\end{equation}
where $J$ is the {\em propagator function} for the Wigner distribution,

\begin{eqnarray}
J(X,K,t;X^{\prime },K^{\prime },0) &=&e^{\gamma t}\ {\frac{1}{{2\pi }}}%
\int_{-\infty }^{\infty }\ dp\ e^{ip[(X-X^{\prime })+\frac{\hbar }{M\gamma }%
(K-K^{\prime })]}\int_{-\infty }^{\infty }\ dx\ e^{-ix(e^{\gamma
t}K-K^{\prime })}  \nonumber \\
&&e^{-D\left( \eta (t)x^{2}-2\zeta (t)\frac{\hbar }{M\gamma }xp+(\frac{\hbar 
}{M\gamma })^{2}tp^{2}\right) }.  \label{A.20}
\end{eqnarray}
We can write this propagator explicitly by preforming the integrations over $%
x$ and $p$, which gives

\begin{equation}
J(X,K,t;X^{\prime },K^{\prime },0)=e^{\gamma t}\ \sqrt{\frac{u}{D\Theta }}%
e^{-u(e^{\gamma t}K-K^{\prime })^{2}}e^{-\Phi /4D\Theta },  \label{A.21}
\end{equation}
where we used the following definitions

\begin{eqnarray}
u(t) &\equiv &1/4D\eta (t)  \nonumber \\
\lambda (t) &\equiv &\frac{\zeta ^{2}(t)}{\eta (t)}=\frac{{2}}{{\ \gamma }}%
\frac{\left( {e^{\gamma t}-1}\right) }{\left( {e^{\gamma t}+1}\right) }, 
\nonumber \\
\vartheta (K,K^{\prime },t) &\equiv &\frac{\hbar }{2M}\lambda
(t)(K+K^{\prime }),  \nonumber \\
\Phi (X,K,X^{\prime },K^{\prime },t) &\equiv &\left[ (X-X^{\prime
})-\vartheta \right] ^{2},  \nonumber \\
\Theta (t) &\equiv &\left( \frac{\hbar }{M\gamma }\right) ^{2}\left[
t-\lambda (t)\right] .  \label{A.22}
\end{eqnarray}

Since we are interested in the density matrix in $(K,p),$ we proceed by
transforming back the Wigner distribution in Eq. (\ref{A.19}) using the
inverse of Eq. (\ref{A.1}), and write

\begin{eqnarray}
\rho (p,K;\sigma ,{\sigma }^{\prime };t) &=&{\frac{1}{{2\pi }}}\int_{-\infty
}^{\infty }\ dX^{\prime }\ \int_{-\infty }^{\infty }\ dK^{\prime }\
W(X^{\prime },K^{\prime };\sigma ,{\sigma }^{\prime };0)\int_{-\infty
}^{\infty }\ dX\ e^{-iXp}J(X,K,t;X^{\prime },K^{\prime },0)\   \nonumber \\
&=&{\frac{1}{{2\pi }}}\int_{-\infty }^{\infty }\ dK^{\prime }\int_{-\infty
}^{\infty }\ dX^{\prime }\ \ \widetilde{J}(p,K,t;X^{\prime },K^{\prime
},0)W(X^{\prime },K^{\prime };\sigma ,{\sigma }^{\prime };0),  \label{A.23}
\end{eqnarray}
where the transform of the propagator is given explicitly, with the help of
Eq.~(\ref{A.21}), by

\begin{eqnarray}
\widetilde{J}(p,K,t;X^{\prime },K^{\prime },0) &=&\int_{-\infty }^{\infty }\
dX\ e^{-iXp}J(X,K,t;X^{\prime },K^{\prime },0)  \nonumber \\
&=&e^{\gamma t}\ \sqrt{4\pi u}e^{-u(e^{\gamma t}K-K^{\prime
})^{2}}e^{-iX^{\prime }p}e^{-D\Theta p^{2}-i\vartheta p}.  \label{A.24}
\end{eqnarray}
Looking at the integral over $X^{\prime }$ in Eq. (\ref{A.23}), we note that
the $X^{\prime }$ dependence of the propagator $\widetilde{J}%
(p,K,t;X^{\prime },K^{\prime },0)$ as it appears in Eq. (\ref{A.24}), is
only through the factor $e^{-iX^{\prime }p}.$ Hence this integral is
actually a Furrier transform of the initial Wigner function, which bring us
back to the initial density matrix

\begin{equation}
\rho (p,K;\sigma ,{\sigma }^{\prime };0)=\int_{-\infty }^{\infty }\
dX^{\prime }\ e^{-iX^{\prime }p}W(X^{\prime },K^{\prime };\sigma ,{\sigma }%
^{\prime };0).  \label{A.25}
\end{equation}
This last result enable us to write the density matrix at any time $t$, in
terms of the density matrix at time $t=0$

\begin{equation}
\rho (p,K;\sigma ,{\sigma }^{\prime };t)={\frac{1}{{2\pi }}}\int_{-\infty
}^{\infty }\ dK^{\prime }\ \widetilde{J}_{\rho }(p,K,t;K^{\prime },0)\rho
(p,K^{\prime };\sigma ,{\sigma }^{\prime };0),  \label{A.26}
\end{equation}
where here the propagator is given simply by Eq. (\ref{A.24}) without the $%
e^{-iX^{\prime }p}$ factor

\begin{equation}
\widetilde{J}_{\rho }(p,K,t;K^{\prime },0)=\ e^{\gamma t}\ \sqrt{4\pi u}%
e^{-u(e^{\gamma t}K-K^{\prime })^{2}}e^{-D\Theta p^{2}-i\vartheta p}.
\label{A.27}
\end{equation}

Consider now the initial condition for the density matrix (\ref{3.33})

\begin{equation}
\rho (p,K;\sigma ,{\sigma }^{\prime },0)=a_{{\sigma }^{\prime
}}^{*}a_{\sigma }2(2\pi {\Delta }^{2})^{1/2}e^{-2{\Delta }%
^{2}K^{2}-iKX_{0}(\sigma -{\sigma }^{\prime })}e^{-{\Delta }%
^{2}p^{2}/2-ipX_{0}(\sigma +{\sigma }^{\prime })/2}.  \label{A.28}
\end{equation}
Substituting Eq.~(\ref{A.28}) into Eq.~(\ref{A.26}), using Eq. (\ref{A.27})
and preforming the integration over $K^{\prime },$ we get the explicit form
of the density matrix in $(p,K)$ space, at time $t,$

\begin{eqnarray}
\rho (p,K;\sigma ,{\sigma }^{\prime };t) &=&a_{{\sigma }^{\prime
}}^{*}a_{\sigma }e^{\gamma t}\ \sqrt{\frac{8\pi u{\Delta }^{2}}{2{\Delta }%
^{2}+u}}e^{-{\Delta }^{2}p^{2}/2-ipX_{0}(\sigma +{\sigma }^{\prime
})/2-D\Theta p^{2}}e^{-i\frac{\hbar }{2M}\lambda Kp-ue^{2\gamma t}K^{2}} 
\nonumber \\
&&\times \exp \left\{ -\frac{\left[ X_{0}(\sigma -{\sigma }^{\prime })+\frac{%
\hbar }{2M}\lambda p+i2ue^{\gamma t}K\right] ^{2}}{4\left( 2{\Delta }%
^{2}+u\right) }\right\} ,  \label{A.29}
\end{eqnarray}
which is our goal in this derivation.

\section{Appendix B}

\subsection{\it The Pointer in a Random Velocity Field}

In Section IV we derived the master equation for the behavior of the pointer
under the influence of the reservoir. Our starting point was the explicit
form of the interaction Hamiltonian $\phi $, which after the averaging over
the bath determined the inverse relaxation time $\gamma .$ As a complement
to this explicit derivation, it will be instructive to look at a much
simpler model, in which the effect of the {\em environment} is described in
terms of a {\em \ random velocity field}, $\ {\bf v}(t).\ $The interaction
of the pointer with the bath, which is simulated by this random field,
perturbs the meter and introduces decoherence. Namely it destroys the
interference by introducing randomization of the phases, and making the
pointer's readings distinguishable. The development in time of the entire
system is described by unitary transformation, which originates probability
into the dynamics by quantum mechanics. This simple model is interesting
because in addition to being readily understood, it gives essentially the
same behavior as the more elaborate treatment of section IV.

Thus taking the Hamiltonian (\ref{2.1}) and adding the random field
interaction we have

\begin{equation}
H=\frac{{\bf P}^{2}}{2M}+V(t)P_{x}\sigma _{z}+{\bf v}(t)\cdot {\bf P}.
\label{B.1}
\end{equation}
The effect of the environment on the pointer is assumed to be represented by
the {\em interaction} Hamiltonian

\begin{equation}
H_{PB}={\bf v}(t)\cdot {\bf P}.  \label{B.2}
\end{equation}
Here ${\bf v}(t)$ is a random vector field, which simulates the {\em random
impacts} suffered by the center of mass of the pointer, while in contact
with the bath around it. The reservoir itself can be thought of as an
ensemble of many external particles, which {\em collide} with the pointer.
The choice of this Hamiltonian can be thought of as an extension of the spin
particle Hamiltonian, $H_{SP}.$ Namely, it is viewed as corresponding to
many individual impacts on the pointer momentum ${\bf P},$ much like the
impact on the center of mass, due to the atomic system, Eq.~(\ref{1.3}). In
our simple model the reservoir is not represented by a Hamiltonian, but
rather by the characterization of the random ''velocity'' field, to be
introduced at the proper stage of the analysis.

This turns Eq.~(\ref{2.2}) into

\begin{eqnarray}
i\hbar {\dot{\Psi}}({\bf R},\sigma ,t) &=&\left( -\frac{\hbar ^{2}}{2M}\ {%
\frac{\partial ^{2}}{{\partial {\bf R}^{2}}}}-i\hbar V(t)\sigma \frac{%
\partial }{\partial X}\right.  \nonumber \\
&&\left. -i\hbar {\bf v}(t)\cdot {\frac{\partial }{{\ \partial {\bf R}}}}%
\right) \Psi ({\bf R},\sigma ,t),  \label{B.3}
\end{eqnarray}
and Eq.~(\ref{2.5}) into

\begin{equation}
i\hbar {\dot{\Psi}}({\bf k},\sigma ,t)=\left( \frac{\hbar ^{2}k^{2}}{2M}%
+\hbar V(t)k_{x}\sigma +\hbar {\bf k}\cdot {\bf v}(t)\right) \Psi ({\bf k}%
,\sigma ,t).  \label{B.4}
\end{equation}

\noindent The solution of this equation is then

\begin{equation}
\Psi ({\bf k},\sigma ,t)=\Psi ({\bf k},\sigma ,0)e^{-i\omega _{k}t-i%
\overline{X}(t)\ k_{x}\sigma -i{\bf k\cdot x(t)}},  \label{B.5}
\end{equation}
where

\begin{equation}
{\bf x}(t)=\int_{0}^{t}\ dt^{\prime }{\bf v}(t^{\prime }),  \label{B.6}
\end{equation}
is also a random field with the dimensions of {\em length}, and the density
matrix is

\begin{eqnarray}
\rho ({\bf k},\sigma ;{\bf k}^{\prime },\sigma ^{\prime };t) &=&\rho ({\bf k}%
,\sigma ;{\bf k}^{\prime },\sigma ^{\prime };0)e^{-i(\omega _{k}-\omega
_{k^{\prime }})t}  \nonumber \\
&&\times e^{-i\overline{X}(t)\ (k_{x}\sigma -k_{x}^{\prime }\sigma ^{\prime
})-i({\bf k}-{\bf k}^{\prime })\cdot {\bf x}(t)}.  \label{B.7}
\end{eqnarray}

Since ${\bf x}(t)$ is a random field it should be expressed in terms of a 
{\em distribution}, $\ W[{\bf x}(t)].\ $ The velocity field is a sum of many
''impulses'', so does the {\em length random field}, ${\bf x}(t)$, namely

\begin{eqnarray}
{\bf v}(t) &=&\sum_{i}{\ {\bf v}_{i}}(t)=\sum_{i}{\ {\bf v}_{i}}f(t-t_{i}) 
\nonumber \\
{\bf x}(t) &=&\sum_{i}{\ {\bf x}_{i}}=\sum_{i}{\ {\bf v}_{i}\tau _{i}}.
\label{B.8}
\end{eqnarray}

\noindent Here ${\bf v}_{i}$ is the ''strength'' of the $i$-th velocity
impulse, $f(t-t_{i})$ samples the $i$-th impulse, whose duration is $\tau
_{i},$ and the summation is carried out over all the impulses which occurred
during the integration time $t$ in Eq.~(\ref{B.6}). Assume that the
probability density of the {\em individual} ${\bf x}_{i}$ is Gaussian,
namely that

\begin{equation}
P({\bf x}_{i})=(2\pi \sigma _{i}^{2})^{-3/2}\exp \left( {-{\bf x}%
_{i}^{2}/2\sigma _{i}^{2}}\right) .  \label{B.9}
\end{equation}

\noindent The single parameter, $\sigma _{i},$ was introduced to represent
the {\em width} of the ${\bf x}_{i}$ distribution, which is also assumed to
be spherically symmetric. The {\em probability density} to find ${\bf x}(t)$%
, at the time $t,$ to be between ${\bf x}$ and ${\bf x}+d{\bf x}$ is then

\begin{equation}
W[{\bf x}(t)]=\prod_{i}\int \ \ d{\bf x}_{i}P({\bf x}_{i})\ \delta [{\bf x}%
(t)-\sum_{j}{\ {\bf x}_{j}}].  \label{B.10}
\end{equation}

\noindent We now transform the $\delta $-function as

\begin{equation}
\delta [{\bf x}(t)-\sum_{j}{\ {\bf x}_{j}}]={\frac{1}{{\ (2\pi )^{3}}}}%
\int_{-\infty }^{+\infty }\ d{\bf q}e^{-i{\bf q}\cdot [{\bf x}(t)-\sum_{j}{\ 
{\bf x}_{j}}]},  \label{B.11}
\end{equation}
and Eq.~(\ref{B.10}) is written as

\begin{eqnarray}
W[{\bf x}(t)] &=&{\frac{1}{{\ (2\pi )^{3}}}}\int_{-\infty }^{+\infty }\ d%
{\bf q}e^{-i{\bf q}\cdot {\bf x}(t)}  \nonumber \\
&&\times \prod_{i}\int \ d{\bf x}_{i}P({\bf x}_{i})\ e^{i{\bf q}\cdot {\bf x}%
_{i}}.  \label{B.12}
\end{eqnarray}
Performing the ${\bf x}_{i}$-integration using Eq.~(\ref{B.9}),

\begin{equation}
\int_{-\infty }^{+\infty }\ d{\bf x}_{i}P({\bf x}_{i})e^{i{\bf q}\cdot {\bf x%
}_{i}}=e^{-\sigma _{i}^{2}q^{2}/2},  \label{B.13}
\end{equation}
the product of the exponentials yields

\begin{equation}
W[{\bf x}(t)]={\frac{1}{{\ (2\pi )^{3}}}}\int_{-\infty }^{+\infty }\ d{\bf q}%
e^{-i{\bf q}\cdot {\bf x}(t)}e^{-(1/2)q^{2}\sum_{i}\sigma _{i}^{2}}.
\label{B.14}
\end{equation}

\noindent Given that the {\em impulses' rate}, i.e., the number of impulses
per unit time is $\nu ,$ and that the {\em average width} of the $\sigma
_{i} $-distribution is $\overline{\sigma },$ we get

\begin{equation}
\sum_{i}{\sigma _{i}^{2}}\to \nu \overline{\sigma }^{2}t,  \label{B.15}
\end{equation}
and finally,

\begin{eqnarray}
W[{\bf x}(t)] &=&{\frac{1}{{\ (2\pi )^{3}}}}\int_{-\infty }^{+\infty }\ d%
{\bf q}e^{-i{\bf q}\cdot {\bf x}(t)}e^{-(1/2)q^{2}\nu \overline{\sigma }%
^{2}t}.  \nonumber \\
&=&\frac{1}{(2\pi \nu \overline{\sigma }^{2}t)^{3/2}}\exp \left( {-}\frac{%
{\bf x}^{2}(t)}{2\nu \overline{\sigma }^{2}t}\right) .  \label{B.16}
\end{eqnarray}
Once the Distribution of ${\bf x}(t)$ is known, we can find the average of
the density matrix in ${\bf k}$-space. Notice that we could first return to
the real space by Eq.~(\ref{2.4}), and then perform the averaging. The
average over the random field is simply

\begin{equation}
\int_{-\infty }^{+\infty }\ d{\bf x}W({\bf x})e^{-i({\bf k}-{\bf k}^{\prime
})\cdot {\bf x}}=e^{-(\beta ^{2}({\bf k}-{\bf k}^{\prime })^{2}/2)}.
\label{B.17}
\end{equation}
where

\begin{equation}
\beta ^{2}=\nu \overline{\sigma }^{2}t.  \label{B.18}
\end{equation}
Notice that Eq.~(\ref{B.17}) is similar to the {\em characteristic function}
of the distribution in ${\bf x.}$

The density matrix of Eq.(\ref{B.7}), averaged over the distribution
function of Eq.(\ref{B.16}), and denoted by ${\overline{\rho }},$ is

\begin{eqnarray}
{\overline{\rho }}({\bf k},\sigma ;{\bf k}^{\prime },\sigma ^{\prime };t)
&=&\rho ({\bf k},\sigma ;{\bf k}^{\prime },\sigma ^{\prime };0)e^{-i(\omega
_{k}-\omega _{k^{\prime }})t}  \nonumber \\
&&\times e^{-i\overline{X}\ (k_{x}\sigma -k_{x}^{\prime }\sigma ^{\prime
})-\beta ^{2}({\bf k}-{\bf k}^{\prime })^{2}/2}.  \label{B.19}
\end{eqnarray}

\noindent We now use Eq.~(\ref{2.10}) and Eq. (\ref{2.4}) to return to real
space and find

\begin{eqnarray}
{\overline{\rho }}({\bf R},\sigma ;{\bf R}^{\prime },\sigma ^{\prime };t)
&=&a_{\sigma ^{\prime }}^{*}a_{\sigma }(8\pi \Delta ^{2})^{3/2}{\frac{1}{{%
(2\pi )^{6}}}}\times  \nonumber \\
&&\int_{-\infty }^{+\infty }\ d^{3}k^{\prime }\ e^{-i{\bf k}^{\prime }\cdot 
{\bf R}^{\prime }}\int_{-\infty }^{+\infty }\ d^{3}k\ e^{i{\bf k\cdot }{\bf R%
}}\ \ e^{-\Delta ^{2}({\bf k}^{2}+{\bf k}^{\prime }{}^{2})}e^{-i(\omega
_{k}-\omega _{k^{\prime }})t-i\overline{{\bf R}}\cdot ({\bf k}\sigma -{\bf k}%
^{\prime }\sigma ^{\prime })-\beta ^{2}({\bf k}-{\bf k}^{\prime })^{2}/2},
\label{B.20}
\end{eqnarray}
where, for convenience of calculation, we have introduced a {\em deflection
vector}

\begin{equation}
\overline{{\bf R}}=(\overline{X},0,0).  \label{B.21}
\end{equation}
We notice that the integral in Eq. (\ref{B.20}) is a product of three
equivalent integrals,

\begin{equation}
{\overline{\rho }}({\bf R},\sigma ;{\bf R}^{\prime },\sigma ^{\prime
};t)=a_{\sigma ^{\prime }}^{*}a_{\sigma }I_{x}I_{y}I_{z},  \label{B.22}
\end{equation}
where, for $i=x,y,z,$

\begin{eqnarray}
I_{i} &=&(8\pi \Delta ^{2})^{1/2}{\frac{1}{{(2\pi )^{2}}}}\int_{-\infty
}^{+\infty }\ dk_{i}^{\prime }\ e^{-ik_{i}^{\prime }R_{i}^{\prime
}}\int_{-\infty }^{+\infty }\ dk_{i}\ e^{ik_{i}R_{i}}e^{-\Delta
^{2}(k_{i}^{2}+k_{i}^{\prime }{}^{2})}e^{-i(\omega _{k_{i}}-\omega
_{k_{i}^{\prime }})t-i{\overline{R}}_{i}\ (k_{i}\sigma -k_{i}^{\prime
}\sigma ^{\prime })-\beta ^{2}(k_{i}-k_{i}^{\prime })^{2}/2}  \nonumber \\
&=&\left( \frac{1}{{2\pi }\Delta _{\beta }^{2}(t)}\right) ^{1/2}\exp \left[ -%
\frac{1}{4\Delta _{\beta }^{2}(t)}\left( \zeta (t)\Omega _{i}^{\prime
2}+\zeta ^{*}(t)\Omega _{i}^{2}+(\Omega _{i}^{\prime }-\Omega
_{i})^{2}(\beta ^{2}/2\Delta ^{2})\right) \right]  \label{B.23}
\end{eqnarray}

\noindent Here we have introduced

\begin{eqnarray}
\Omega _{i} &=&R_{i}-{\overline{R}}_{i}\sigma ,  \nonumber \\
\Omega _{i}^{\prime } &=&R_{i}^{\prime }-{\overline{R}}_{i}\sigma ^{\prime }
\nonumber \\
\Delta _{\beta }^{2}(t) &=&\Delta ^{2}[\xi (t){+}(\beta ^{2}/\Delta ^{2})],
\label{B.24}
\end{eqnarray}
where in the last equation, $\Delta _{\beta }^{2}(t)$ reflects the spatial
broadening of the pointer's position. This is due to both the free quantum
diffusion, and an additional diffusion term, $\beta ^{2},$ caused by the
random field ${\bf v}(t),$ which simulates the environment. Using Eq.~(\ref
{B.23}) we can write the three dimensional density matrix of Eq.~(\ref{B.20}%
) as

\begin{eqnarray}
{\overline{\rho }}({\bf R},\sigma ;{\bf R}^{\prime },\sigma ^{\prime };t)
&=&a_{\sigma ^{\prime }}^{*}a_{\sigma }\left( \frac{1}{{2\pi }\Delta _{\beta
}^{2}(t)}\right) ^{3/2}  \nonumber \\
&&\times \exp \left[ -\frac{1}{4\Delta _{\beta }^{2}(t)}\left( \zeta (t){\bf %
\Omega }^{\prime 2}+\zeta ^{*}(t){\bf \Omega }^{2}+({\bf \Omega }^{\prime }-%
{\bf \Omega })^{2}\frac{\beta ^{2}}{2\Delta ^{2}}\right) \right] ,
\label{B.25}
\end{eqnarray}
where we now have

\begin{eqnarray}
{\bf \Omega } &\equiv &{\bf R}-\widehat{{\bf x}}\overline{X}\sigma , 
\nonumber \\
{\bf \Omega }^{\prime } &\equiv &{\bf R}^{\prime }-\widehat{{\bf x}}%
\overline{X}\sigma ^{\prime }.  \label{B.26}
\end{eqnarray}
Eq.~(\ref{B.25}) is our result for the reduced density matrix of the pointer
in space, after the environment was {\em ''traced out''}. Notice that the
spin indices can be considered as parameters for the pointer's position. It
is clearly seen that when $\beta \rightarrow 0$ Eq.~(\ref{B.25}) is reduced
back to Eq.~(\ref{2.11}).

We now turn to calculate the probability to locate the pointer in the
position ${\bf R}$ in space. We set ${\bf R}^{\prime }={\bf R}$ in Eq.(\ref
{B.25}), and find the diagonal elements in space of the density matrix, ${%
\overline{\rho }}({\bf R},\sigma ;{\bf R},\sigma ^{\prime };t).$ Again we
write this spatial probability density, as in Eq.~(\ref{2.17}), as a sum of
three terms. The first term is correlated with the {\em spin up} state, and
we set $\sigma ^{\prime }=\sigma =+1,$ or,

\begin{equation}
{\bf \Omega }^{\prime }={\bf \Omega }={\bf R}-\hat{{\bf x}}\overline{X},
\label{B.27}
\end{equation}
which leads to: 
\begin{eqnarray}
P_{+1}({\bf R},t) &=&\mid a_{+1}\mid ^{2}\left( \frac{1}{{2\pi }\Delta
_{\beta }^{2}(t)}\right) ^{3/2}  \nonumber \\
&&\times \exp \left( -{\frac{1}{{2}\Delta _{\beta }^{2}(t)}[}{\bf R}-\hat{%
{\bf x}}\overline{X}]^{2}\right) {.}  \label{B.28}
\end{eqnarray}
The second term is correlated with the {\em spin down} state, and we set $%
\sigma ^{\prime }=\sigma =-1,$ or,

\begin{equation}
{\bf \Omega }^{\prime }={\bf \Omega }={\bf R}+\hat{{\bf x}}\overline{X},
\label{B.29}
\end{equation}
and find

\begin{eqnarray}
P_{-1}({\bf R},t) &=&\mid a_{-1}\mid ^{2}\left( \frac{1}{{2\pi }\Delta
_{\beta }^{2}(t)}\right) ^{3/2}  \nonumber \\
&&\times \exp \left( -{\frac{1}{{2}\Delta _{\beta }^{2}(t)}[}{\bf R}+\hat{%
{\bf x}}\overline{X}]^{2}\right) {.}  \label{B.30}
\end{eqnarray}

The probabilities of Eqs. (\ref{B.28}) and (\ref{B.30}) are similar to these
of Eqs.~(\ref{2.18a}) and (\ref{2.18b}). The only difference is that the 
{\em free diffusion} spread of the Gaussians, $\Delta _{f}^{2}(t),$ is
replaced by $\Delta _{\beta }^{2}(t)$ of Eq.~(\ref{B.24}), which adds the
effect of the environment on the pointer's position. We thus observe that
the probability to find the pointer, either deflected by $+\overline{X}$ \
for the spin up case, or by $-\overline{X}$ \ for the spin down case, is
''performing'' a diffusion-like motion in space, which is strongly affected
by the random impulses inflicted by the environment. It is interesting to
notice that even in the absence of interaction with the spin particle (i.e.
if $\overline{X}=0$), the pointer's position would have shown diffusion-like
motion of its Gaussian wave-packet, with the same time dependent spread, $%
\Delta _{\beta }^{2}(t)$.

This diffusion-like behavior of the pointer, due to the interaction with the
environment, is a reminiscent of the classical diffusion of a Brownian
particle immersed in a fluid \cite{Chandrasekhar 43}. If we compare the
result of Eq.~(\ref{B.16}) with the analogous one of the Brownian motion, we
find that the {\em diffusion coefficient} is $D=\beta ^{2}/(2t)$ or with
Eq.~(\ref{B.18}) $D=\nu {\overline{\sigma }}^{2}/2.$ This can provide us
with an estimate for ${\overline{\sigma }}.$ Assuming that the environment
is a bath in thermal equilibrium with temperature $T,$ then, in analogy with
Brownian particle, the classical limit yields ${\overline{\sigma }}^{2}~\to
~2k_{B}T/(M\nu ^{2}),$ where $k_{B}$ is Boltzmann's constant. If we consider
the environment as an ensemble of harmonic oscillators, with a typical
frequency of $\omega _{0},$ we get

\[
{\overline{\sigma }}^{2}\to \left( \frac{2}{M\nu ^{2}}\right) {\frac{{\
\hbar \omega _{0}}}{{\ e^{(\hbar \omega _{0}/k_{B}T)}-1}}}, 
\]

\noindent which approaches the classical limit when $(\hbar \omega
_{0}/k_{B}T)~\ll ~1.$ The collision rate, $\nu ,$ which is the inverse {\em %
relaxation time}, $\tau _{r},$ of the pointer, depends of course on the
character of the bath.

It is also instructive to estimate the time scale, on which this diffusion
broadening becomes so wide, as to erase the record of the measurement. We
can estimate this time, $t_{bluer},$ by noting that when the width $\Delta
_{\beta }(t_{bluer})$ is comparable with the deflection of the pointer $%
\overline{X}$, it becomes difficult to distinguish between the spin up and
spin down results, which were previously registered. Thus taking $\overline{X%
}^{2}/\Delta ^{2}\gg 1,$ which is a necessary condition for a meaningful
pointer, and $\xi (t)$ $\simeq 1$, we have

\begin{equation}
t_{bluer}\simeq \tau _{r}(\overline{X}/\overline{\sigma })^{2}.  \label{B.31}
\end{equation}
This ''theoretical'' time of {\em bluer} is a consequence of the {\em %
classical} Brownian-like diffusion of the pointer, and since always $%
\overline{X}\gg $ $\overline{\sigma },$ this time is much greater then the
relaxation time.

We turn, now, to the off-diagonal term of the spatial probability of the
pointer, namely to the {\em interference} term, which is a purely {\em %
quantum mechanical }effect, like that of Eq.~(\ref{2.19}). We set, again, in
Eq.~(\ref{B.25}), ${\bf R}^{\prime }={\bf R},$ but now take $\sigma ^{\prime
}=-1$, and $\sigma =+1,$ and have

\begin{eqnarray}
{\bf \Omega }^{^{\prime }} &=&{\bf R}+\widehat{{\bf x}}\overline{X}, 
\nonumber \\
{\bf \Omega } &=&{\bf R}-\widehat{{\bf x}}\overline{X},  \label{B.32}
\end{eqnarray}
thus using Eqs.~(\ref{B.28}) and (\ref{B.30}) we get an expression similar
to that of Eq.~(\ref{2.19}), i.e.,

\begin{equation}
P_{-1,+1}({\bf R},t)=2\sqrt{P_{+}({\bf R},t)P_{-}({\bf R},t)}\cos \left( 
\frac{{\bf R\cdot }\overline{{\bf X}}}{\Delta _{\beta }^{2}(t)}\frac{t}{\tau
_{f}}+\varphi _{-}-\varphi _{+}\right) \exp \left\{ -\frac{\overline{X}%
^{2}(\beta ^{2}/\Delta ^{2})}{2\Delta _{\beta }^{2}(t)}\right\} ,
\label{B.33}
\end{equation}

Comparing this interference probability with that of the {\em free} pointer,
Eq.~(\ref{2.19}), we notice that in Eq.~(\ref{B.33}), beside the replacement
of $\Delta _{f}^{2}(t)$ by $\Delta _{\beta }^{2}(t),$ an additional
exponential factor, $e^{-g(t)}$, appears, with

\begin{equation}
g(t)=\frac{\overline{X}^{2}(\beta ^{2}/\Delta ^{2})}{2\Delta _{\beta }^{2}(t)%
}.  \label{B.34}
\end{equation}
This damping effect stems, uniquely, from the interaction with the bath, and
it is similar to that found in Refs. \cite{Caldeira 85,Milburn 85}. This
exponential factor will kill the interference term almost instantaneously.
To show it, we rewrite the {\em damping function}, $g(t),$ of Eq.~(\ref{B.33}%
) as

\begin{equation}
g(t)={\frac{1}{2}}\left( \frac{\overline{X}}{\Delta }\right) ^{2}\ {\frac{{\
({\frac{{\overline{\sigma }}}{{\Delta }}})^{2}{\frac{t}{{\tau _{r}}}}}}{{\ 1+%
{\ ({\frac{{\overline{\sigma }}}{{\Delta }}})^{2}{\frac{t}{{\tau _{r}}}}}}}},
\label{B.35}
\end{equation}
where we took $\xi (t)=1.$ First we investigate $g(t),$ at early times, when 
$t\ll \tau _{r}({\frac{{\Delta }}{{\overline{\sigma }}}})^{2}.$ We see that $%
g(t)\to t/\tau _{int\ },$ where the {\em interference damping time} is given
by

\begin{equation}
\tau _{int\ }=2\tau _{r}\left( \frac{\Delta }{\overline{X}}\right)
^{2}\left( {\frac{{\Delta }}{{\overline{\sigma }}}}\right) ^{2},
\label{B.36}
\end{equation}
which is indeed extremely small compared to $\tau _{r}.$ To make a simple
estimate let us take the relevant parameters for a silver atom pointer i.e. $%
{\Delta \sim 1}$ ${\rm \mu m,}$ ${\overline{\sigma }\sim 0.1}$ ${\rm \mu m}$
and $\overline{X}\sim 1$ {\rm cm}.This yields $\tau _{int\ }/\tau _{r}\simeq
10^{-10}.$ It is interesting to compare Eq. (\ref{B.35}) with Eq. (\ref{3.52}%
) of the gas reservoir model. We see that both expressions are formally
identical. Taking $D=\gamma Mk_{B}T/\hbar ^{2}$ and $\tau _{f}=2M\Delta ^{2}{%
/\hbar }$ we can write the term containing $t$ in the nominator of Eq. (\ref
{3.52}) as

\begin{equation}
2D\left( \frac{2{\Delta }}{\gamma \tau _{f}}\right) ^{2}t=2\left( \frac{{1}}{%
\Delta }\right) ^{2}\frac{k_{B}T}{M\gamma ^{2}}\gamma t.  \label{B.37}
\end{equation}
The analogous term in Eq. (\ref{B.35}) gives

\begin{equation}
\left( \frac{{1}}{{\Delta }}\right) ^{2}\frac{{\overline{\sigma }}^{2}}{\tau
_{r}}t=2\left( \frac{{1}}{{\Delta }}\right) ^{2}\frac{k_{B}T}{M\gamma ^{2}}%
\gamma t,  \label{B.38}
\end{equation}
where we have taken ${\overline{\sigma }}^{2}~\rightarrow ~2k_{B}T/(M\nu
^{2})$, and $\nu =\gamma =1/\tau _{r}.$ We notice that in the regime where $%
g(t)$ is linear in time both the Fokker - Planck and the random field models
give the same behavior. Of course when the time $t$ increases, so does $g(t)$%
, and eventually it reaches a{\em \ saturation} value of $g(t)\rightarrow {%
\frac{1}{2}}\left( \overline{X}/\Delta \right) ^{2}.$

\newpage

\section{\bf Figure Captions}

\vskip 5mm

{\bf Fig. 1}

A graphical description of the decoherence function, $g(t),$ in the exponent
of Eq. (\ref{3.44}), normalized to the saturation value ${\frac{1}{2}}\left( 
\overline{X}/\Delta \right) ^{2}$ and as a function of $\gamma t.$ In the
inset, we see more closely the behavior at initial times, where $g(t)$
starts as $(\Gamma ^{\prime }t)^{3}$ and quickly goes to the linear behavior
as $\Gamma t.$

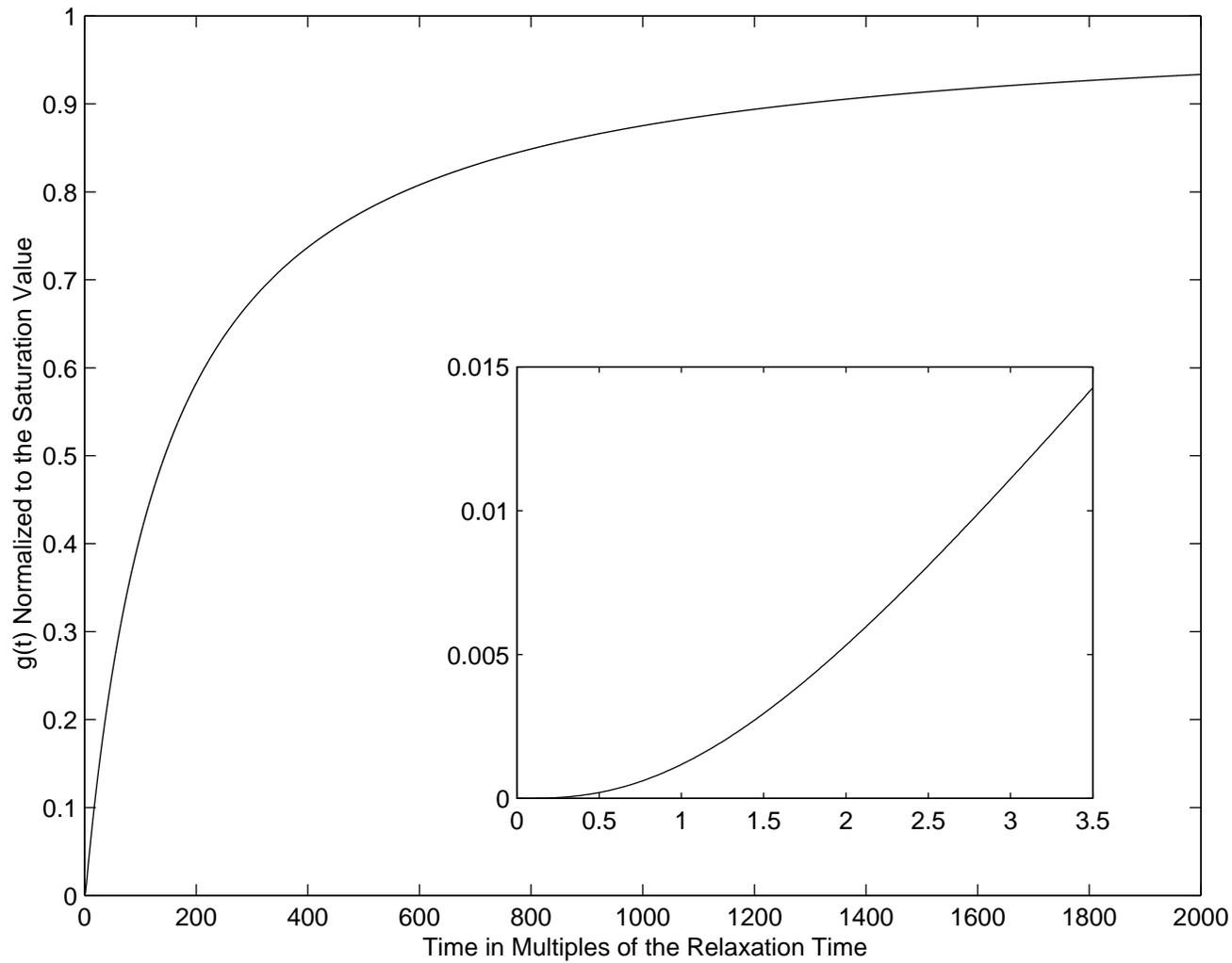
\end{document}